\documentclass{emulateapj}
\usepackage{ulem}
\usepackage[usenames]{color}
\usepackage{graphicx}
\slugcomment{~~\today}
\slugcomment{Accepted for publication in The Astrophysical Journal: \today}

\newcommand       \msun        	{$M_{\odot}$}

\newcommand	     \cc             {cm$^{-3}$}

\newcommand       \mic        	 {$\mu$m}

\newcommand		\irx 	{$IRX$}

\newcommand	\spitz		{{\it Spitzer}}
\newcommand	\chan		{{\it Chandra}}

\begin{document}
\bibliographystyle{/Users/edwek/Library/texmf/tex/latex/misc/aastex52/aas.bst}

\title{Five Years of Mid-Infrared Evolution of the Remnant \\of SN 1987A: The Encounter Between the Blast Wave \\ and the Dusty Equatorial Ring}
\author{Eli Dwek\altaffilmark{1}, Richard G. Arendt\altaffilmark{2}, Patrice Bouchet\altaffilmark{3}, David N. Burrows\altaffilmark{4}, Peter Challis\altaffilmark{5}, I. John Danziger\altaffilmark{6},    James M.~De
Buizer\altaffilmark{7}, Robert D. Gehrz\altaffilmark{8}, Sangwook Park\altaffilmark{4}, Elisha F. Polomski\altaffilmark{9}, Jonathan D. Slavin\altaffilmark{5}, and Charles E. Woodward\altaffilmark{8}}

\altaffiltext{1}{Observational Cosmology Lab., Code 665; NASA Goddard
Space Flight Center, Greenbelt, MD 20771, U.S.A., e-mail: eli.dwek@nasa.gov}
\altaffiltext{2}{CRESST/UMBC, Code 665, NASA Goddard Space Flight Center, Greenbelt MD, 20771, U.S.A.}  
\altaffiltext{3}{DSM/DAPNIA/Service d'Astrophysique, CEA/Saclay, F-91191 Gif-sur-Yvette; Patrice.Bouchet@cea.fr}
\altaffiltext{4}{Department of Astronomy and Astrophysics, Pennsylvania State University, 525 Davey
Laboratory, University Park, PA 16802, U.S.A.} 
\altaffiltext{5}{Harvard-Smithsonian, CfA, 60 Garden St., MS-19, Cambridge, MA 02138, U.S.A.}
\altaffiltext{6}{Osservatorio Astronomico di Trieste, Via Tiepolo, 11,
Trieste, Italy} 
\altaffiltext{7}{Gemini Observatory, Southern
Operations Center, c/o AURA, Casilla 603, La Serena, Chile}
\altaffiltext{8}{Department of Astronomy, University of Minnesota, 116 Church St., SE, Minneapolis, MN 55455, U.S.A.}
\altaffiltext{9}{University of Wisconsin, Eau Claire, WI 54702, U.S.A.}

\begin{abstract}
We have used the {\it Spitzer} satellite to monitor the mid-IR evolution of SN 1987A over a 5 year 
period spanning the epochs between days $\sim$~6000 and 8000 since the explosion. The supernova (SN) has evolved into a supernova remnant (SNR) and its radiative output is dominated by the interaction of the SN blast wave with the pre-existing equatorial ring (ER). The mid-IR spectrum is dominated by emission from $\sim 180$~K silicate dust, collisionally-heated by the hot X-ray emitting gas with a temperature and density of $\sim 5\times10^6$~K and $\sim 3\times 10^4$~\cc, respectively. The mass of the radiating dust is $\sim 1.2\times 10^{-6}$~\msun\ on day 7554, and scales linearly with IR flux. Comparison of the IR data with the soft X-ray flux derived from \chan\ observations shows that the IR-to-Xray flux ratio, \irx, is roughly constant with a value of 2.5. Gas-grain collisions therefore dominate the cooling of the shocked gas. The constancy of \irx\ is most consistent with the scenario that very little grain processing or gas cooling have occurred throughout this epoch. The shape of the dust spectrum remained unchanged during the observations while the total flux increased by a factor of $\sim 5$ with a time dependence of $t'^{0.87\pm0.20}$, $t'$ being the time since the first encounter between the blast wave and the ER. These observations are consistent with the transitioning of the blast wave from free expansion to a Sedov phase as it propagates into the main body of the ER, as also suggested by X-ray observations. The constant spectral shape of the IR emission provides strong constraints on the density and temperature of the shocked gas in which the interaction takes place. Silicate grains, with radii of $\sim 0.2$~\mic\ and temperature of $T\sim 180$~K, best fit the spectral and temporal evolution of the $\sim 8 - 30$~\mic\ data. The IR spectra also shows the presence of a secondary population of very small, hot ($T \gtrsim 350$~K), featureless dust. If these grains spatially coexist with the silicates, then  they must have shorter lifetimes. The data show slightly different rates of increase of their respective fluxes, lending some support to this hypothesis. However, the origin of this emission component and the exact nature of its relation to the silicate emission is still a major unsolved puzzle. 
\end{abstract}

\section{INTRODUCTION}
About 10 years after its explosion on February 23, 1987, supernova (SN) 1987A has evolved from a supernova, when its radiative output was dominated by the release of radioactive decay energy in the ejecta, into a supernova remnant (SNR), when its radiative output became dominated by the interaction of its blast wave with the inner equatorial ring (ER). The ER is located at a distance of about 0.7 lyr from the center of the explosion, and could have been produced by mass loss from a single rotating supergiant  \citep{heger98} or by a merger event in a binary system that also formed the two outer rings \citep{morris09}.
The transition from SN to SNR was observed at wavelengths ranging from radio to X-rays, and is summarized in Figure~18 in \cite{bouchet06}.

In this paper we report on the continuing evolution of the $\sim 5-30$~\mic\ spectrum and the 3.6, 4.5, 5.8, 8.0, and 24~\mic\ photometric fluxes from SN 1987A, spanning the $\sim 5$ year period from day $\sim$~6000 until day $\sim$~8000 after the explosion. Initial reports and analysis of the IR evolution were presented by \cite{bouchet04, bouchet06} and \cite{dwek08a}. In \S2 we present the IR data obtained by the \spitz\ satellite. The evolution of the IR emission and the dust composition are presented in \S3. In \S4 we derive the plasma conditions from the IR observations. The comparison of the IR emission with the X-ray emission, and the evolution of their flux ratio are discussed in \S5. A brief summary of the paper is presented in \S6.

\begin{figure*}[ht!] 
  \plotone{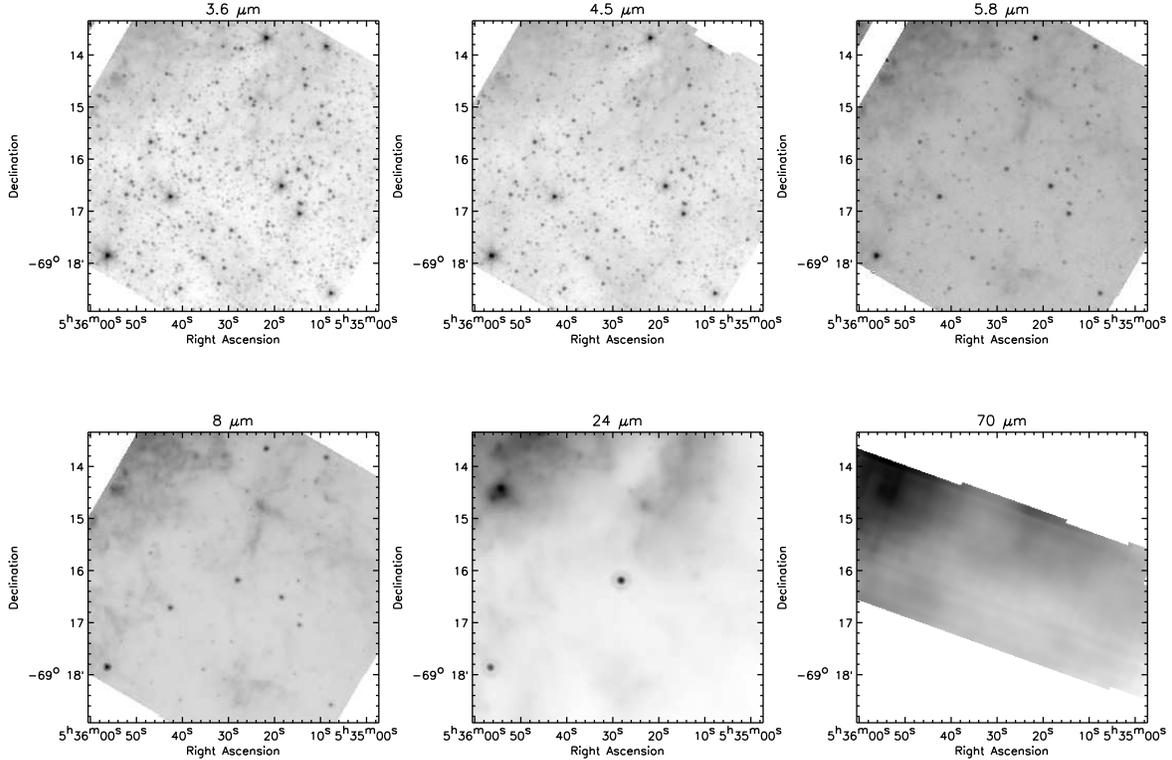}
  \caption{\footnotesize{{\it Spitzer} images of SN 1987A at Day 7975 (3.6 - 8~\mic) and 
  Day 7983 (24 - 70~\mic). These reverse grayscale images use logarithmic scaling.}}
  \label{grey}
\end{figure*} 

\begin{figure*}[ht!] 
  \plotone{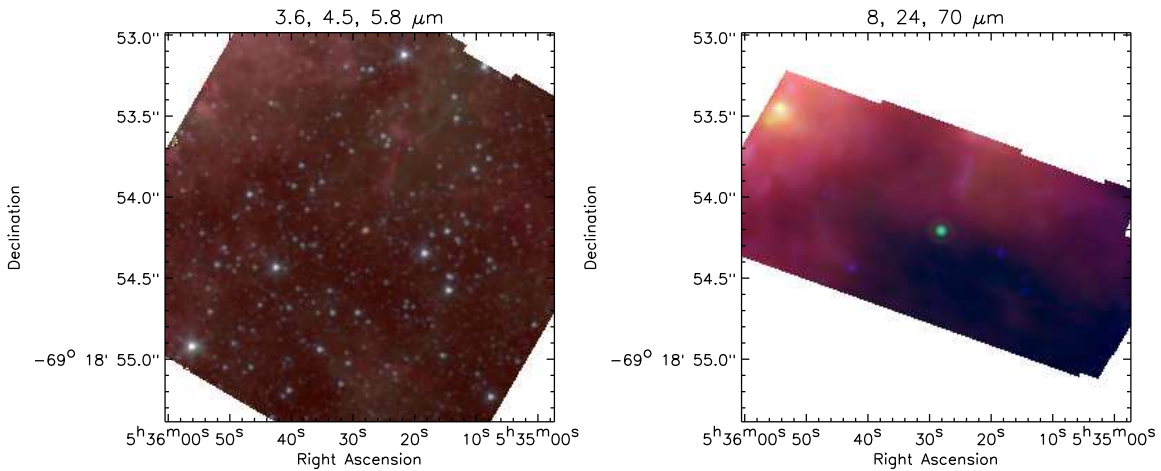}
  \caption{\footnotesize{{\it Spitzer} images of SN 1987A at Day 7975 (3.6 - 8~\mic) and 
  Day 7983 (24 - 70~\mic).}}
  \label{color}
\end{figure*}

\section{OBSERVATIONS AND DATA REDUCTION}
{\it Spitzer's} observations using the InfraRed Array Camera (IRAC), the InfraRed Spectrograph (IRS), and Multiband Imaging Photometer for Spitzer (MIPS) were conducted annually during 
the first two years (2004--2005), and then roughly every 6 months 
for the remainder of its cryogenic mission. SN 1987A was also observed incidentally
with IRAC and MIPS by the SAGE survey \citep{meixner06} and other projects during 2005.

\begin{figure*}[ht] 
\centering
  \includegraphics[width=3.0in]{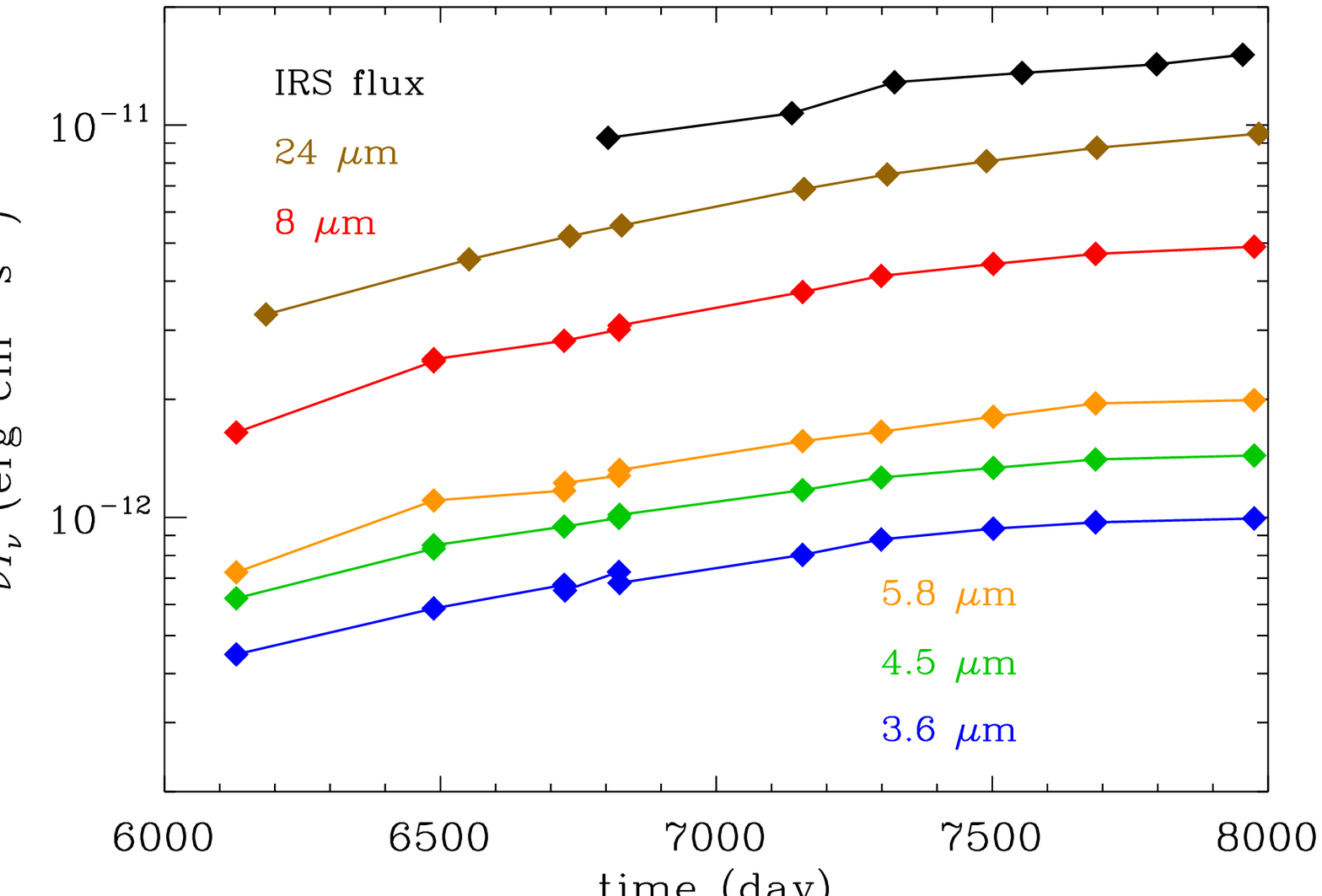}
  \includegraphics[width=3.0in]{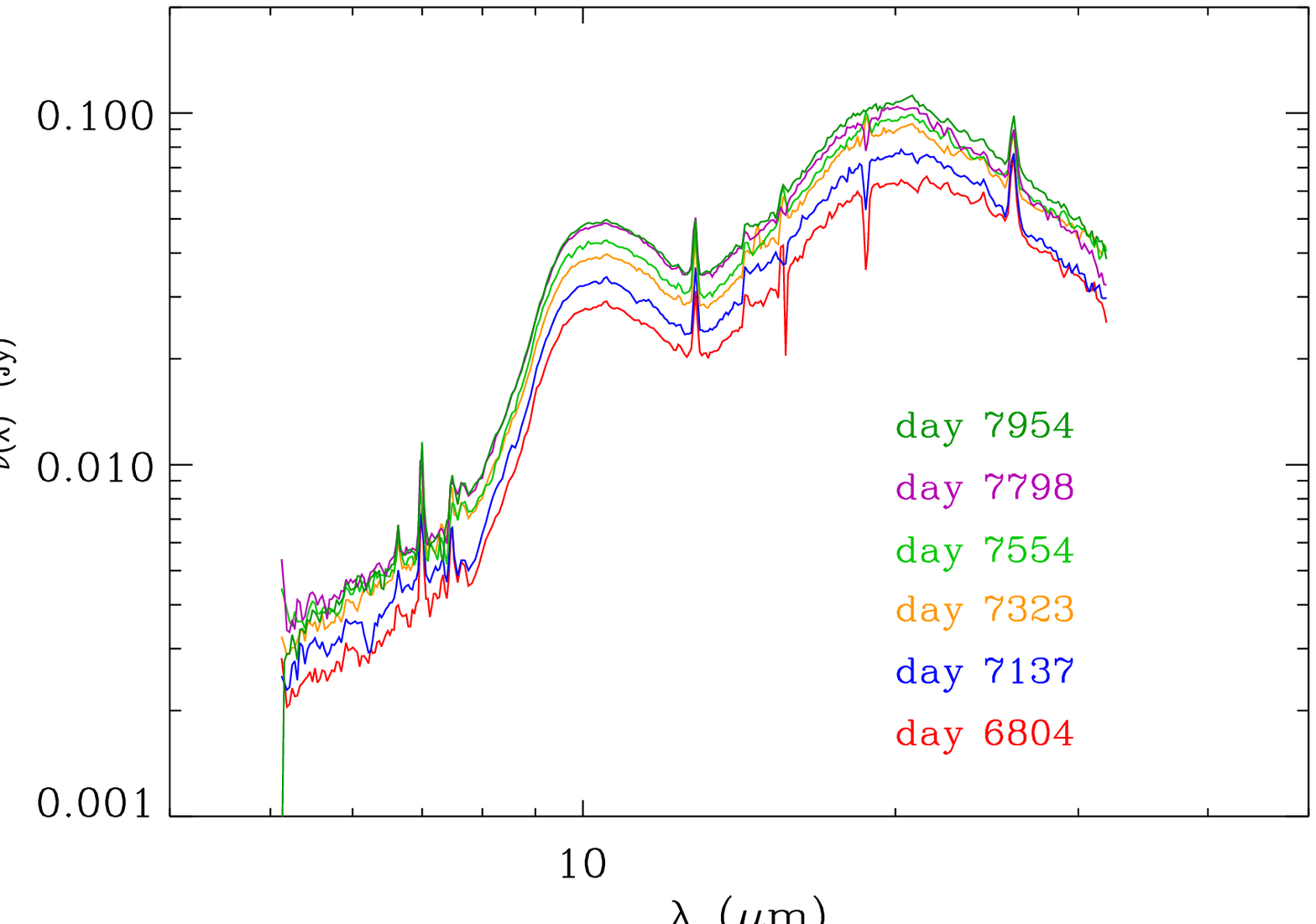} \\
    \includegraphics[width=3.0in]{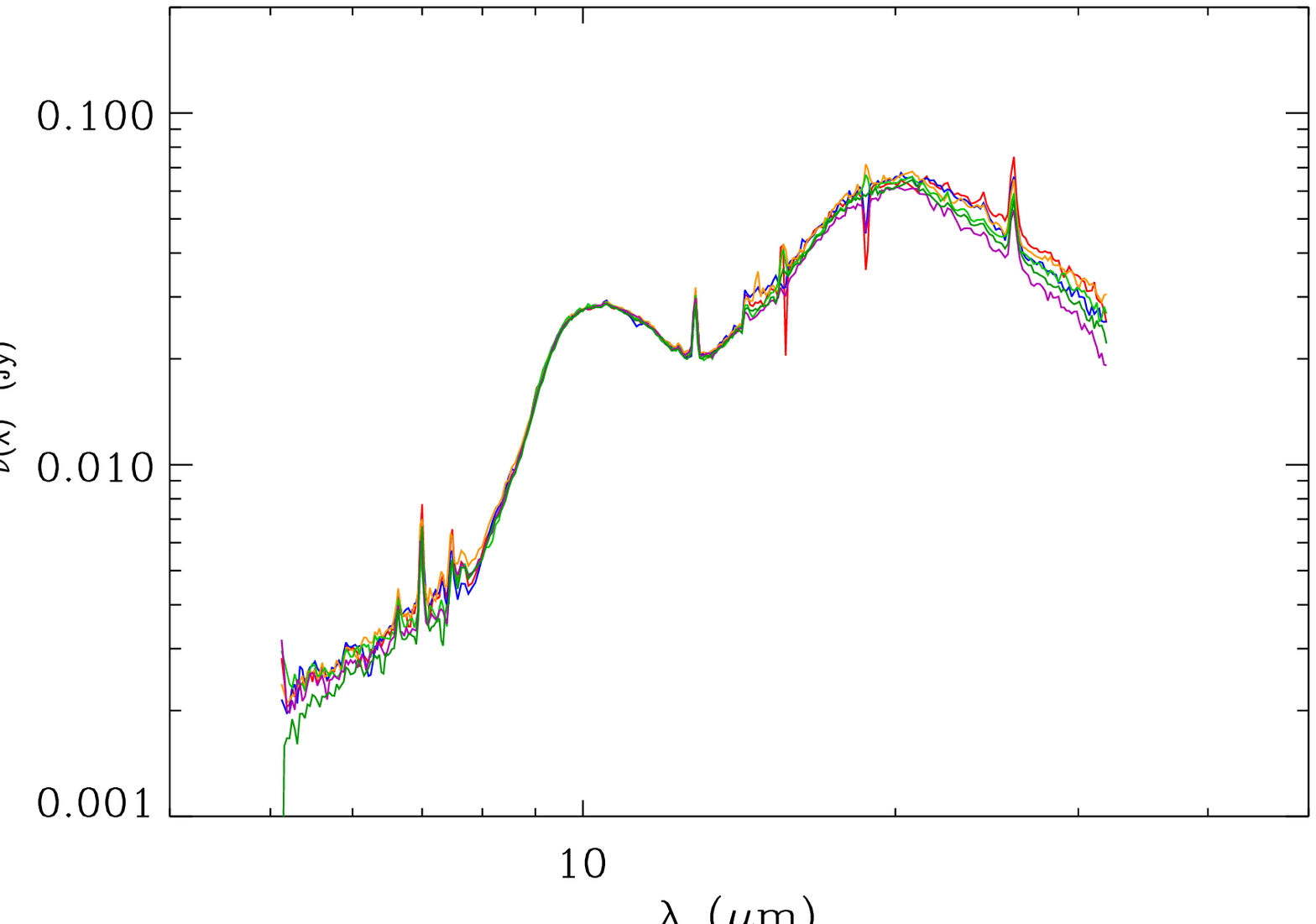}
     \includegraphics[width=3.0in]{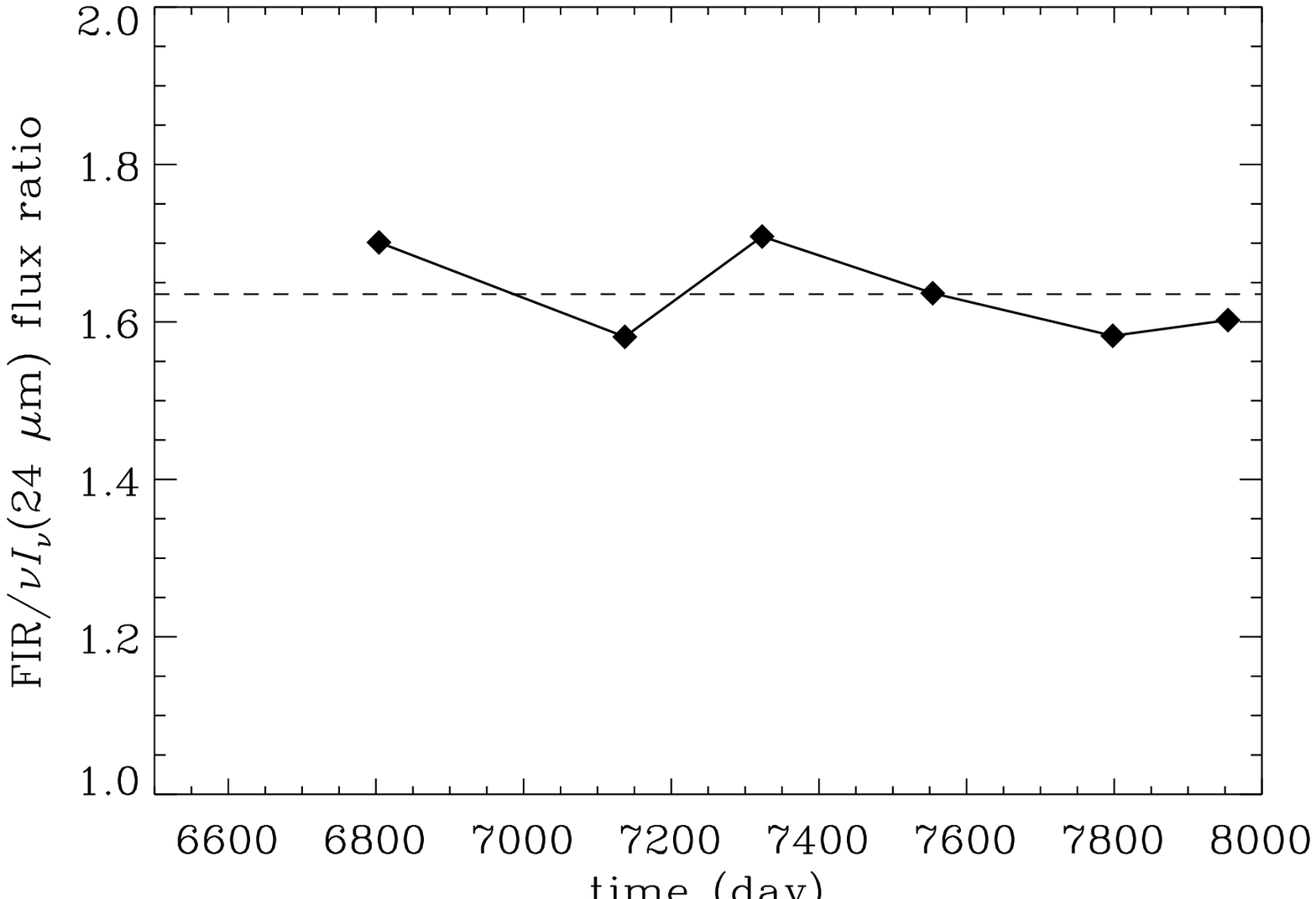}
  \caption{\footnotesize{{\bf Top row}: The evolution of the IRAC and MIPS 24~\mic\ luminosities (left panel) and the specific intensity of the low-resolution spectra (right panel) of the equatorial ring. {\bf Bottom row}: Spectra normalized to the one obtained on day 7554 (left panel). Because of the constancy in the shape of the spectrum, any of the IRAC or MIPS photometric fluxes can be used as a proxy for the integrated IR spectrum from the ER. The right panel depicts the almost constant ratio between the IRS integrated over the 5-30~\mic\ spectrum to the 24~\mic\ luminosity.}}
  \label{IRspec_vol}
\end{figure*} 

Figures \ref{grey} and \ref{color} 
illustrate the IRAC data at the final observations ($\sim$ day 8000).
The IRAC data are all of good quality, i.e., the data are not adversely affected by any 
artifacts induced by bright sources in the field or in prior observations. 
Initial tests of generating mosaics from the basic calibrated data (BCD) showed no significant difference
from the post-BCD mosaics. Therefore, aperture photometry was performed on 
the post-BCD mosaics in order to obtain the broad band evolution
of SN 1987A at 3.6, 4.5, 5.8 and 8~\mic. These measurements and their statistical
uncertainties are shown in Figure \ref{IRspec_vol}. 

\begin{figure*}[ht] 
 \centering
 \includegraphics[width=3.0in]{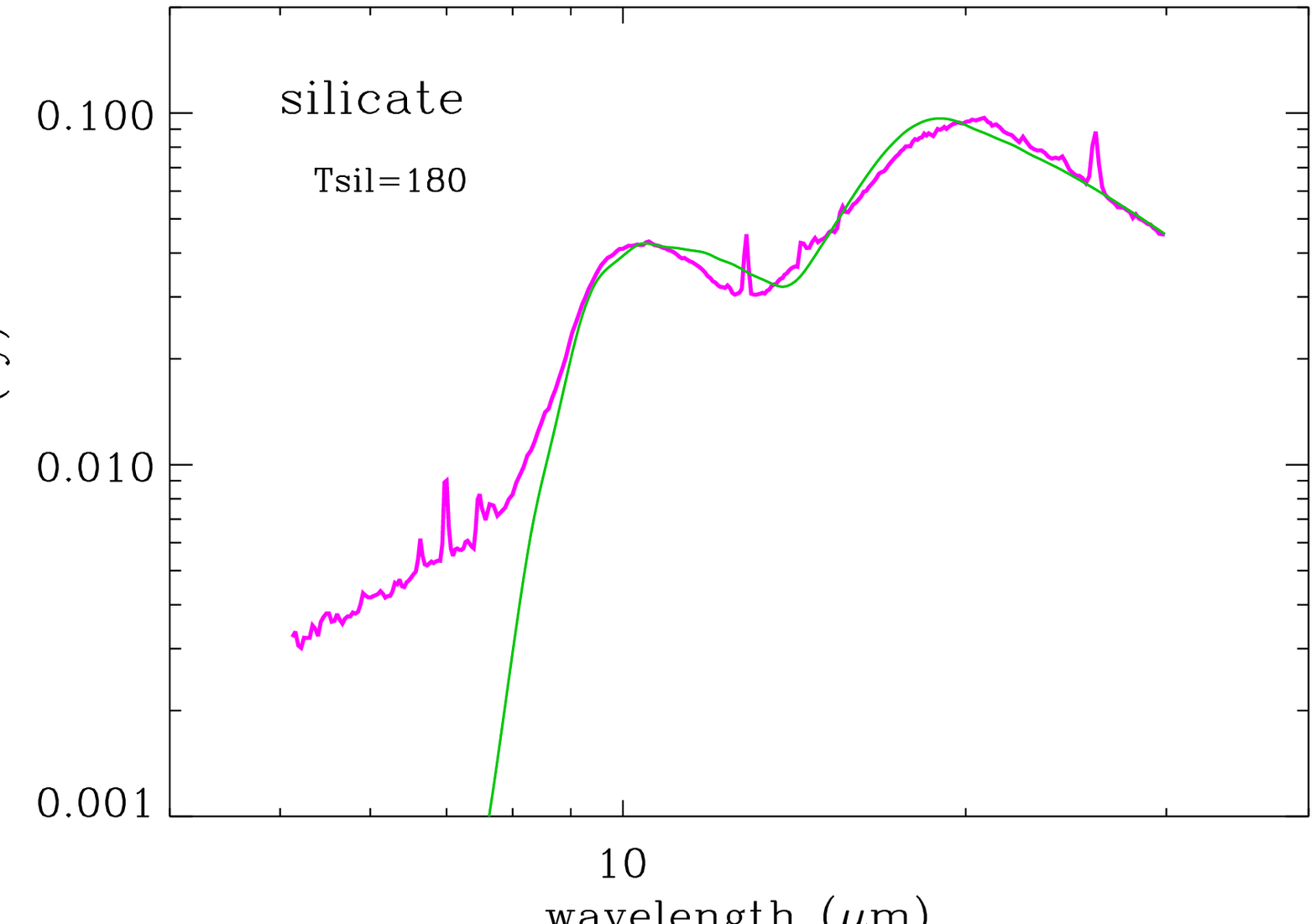} 
    \includegraphics[width=3.0in]{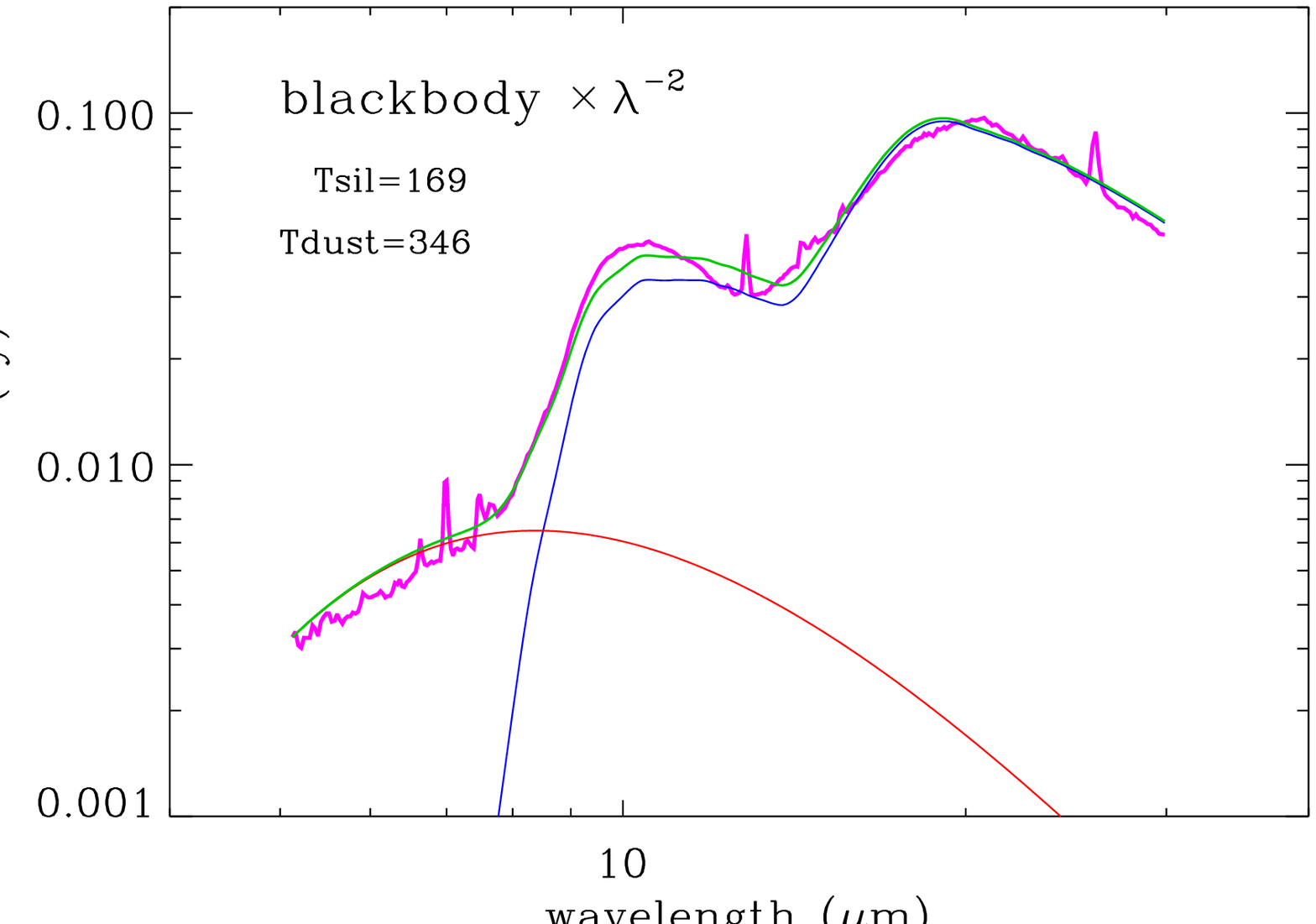} 
  \caption{\footnotesize{{\bf Left panel}: Single composition fit to the IR spectrum of the ER. The fit corresponds to the spectrum of astronomical silicate grains radiating at 180~K. The excess $5-8$~\mic\ emission evolves at the same rate as the silicate emission (see Fig. \ref{IRspec_vol}) strongly suggesting that it is generated by an additional dust component in the ER. {\bf Right panel}: The addition of any dust component with a smooth spectrum which will not reduce the goodness-of-fit to the 9.7 and 18~\mic\ silicate features is a viable candidate for this second dust component. Here the spectrum is fit with a modified blackbody with a $\lambda^{-2}$ emissivity law and a temperature of 346~K.}}
  \label{sil_bb}
\end{figure*} 

Figures \ref{grey} and \ref{color} 
 also illustrate the MIPS 24 and 70~\mic\ images of SN 1987A at Day 7983.
There is no evident point source at 70~\mic. As for IRAC, there seemed to be no
advantage to reprocessing the BCD, thus aperture photometry was performed on the 
post-BCD mosaics. The 24~\mic\ flux densities are shown in Figure \ref{IRspec_vol}. 
At Day 7983, a 
70~\mic\ upper limit was determined by adding an artificial point source
to the MIPS image at the location of the SN. We started with a source flux that was 
clearly visible above the background confusion, and then reduced the flux until the source
was no longer detected using the FIND routine of the IDLASTRO library. 
The results indicate the $S(70 \micron) < 0.09$ Jy.

A significant complication with the IRAC and MIPS photometry is that the SN is not resolved 
from the nearby Star 2 and Star 3. If the stars are assumed to have mid-IR magnitudes 
that are equal to the K magnitudes (15.06, 15.80 respectively from Walborn et al 1993),
then their combined IR flux densities should be 0.41, 0.26, 0.16, 0.09, and 0.01 mJy 
at 3.6, 4.5 5.8, 8, and 24~\mic, respectively. These assumptions may be accurate for Star 2, which 
is identified as a B2 III star \citep{walborn93,scuderi96}. However 
\cite{walborn93} report both, an IR excess and a variability by at least $\Delta m = 1$ 
mag for Star 3. The fluxes in Figure \ref{IRspec_vol} are plotted after subtraction of the emission of Stars 2 and 3 under these assumptions.

IRS observations were reduced starting with the BCD, and using the basic steps outlined in 
the IRS Data Handbook. For the low resolution 
($R \sim 60 -120$) data, we constructed a 2-D background frame
using the observations of a specific off--source target position. IRSCLEAN$\_$MASK 
was used to identify and clean rouge pixels in the background--subtracted BCD frames. 
The cleaned BCD data were averaged for each nod (positions of the source at 1/3 or 2/3 
the length of the slit), and spectra were extracted from these super BCDs 
using the Spitzer IRS Custom Extraction (SPICE) software\footnote{\url{http://ssc.spitzer.caltech.edu/dataanalysistools/tools/spice/}}. After extraction, the spectra orders were merged to produce the full 
spectra shown in Figure \ref{IRspec_vol} (top right panel). The high resolution ($R \sim 600$) 
data were processed in a similar fashion. The major difference was that instead of 
using a background from a dedicated off-source target, the background is derived from the 
average of two pointings that flank the SN on opposite sides ($15''$ and $25''$ away for
the short high resolution (SH) and long high resolution (LH) spectra respectively). The position angle of these 
pointings varied as dictated by the scheduling of the observations, but the distance of the 
pointings from the SN remained constant. For both the high and low resolution data, we used the Spectroscopic Modeling Analysis and Reduction Tool\footnote{\url{http://ssc.spitzer.caltech.edu/dataanalysistools/tools/contributed/irs/smart/}}
 (SMART) to fit and extract line fluxes.

\begin{deluxetable*}{lccc}
\centering
\tabletypesize{\footnotesize}
\tablewidth{0pt}
\tablecaption{Observed X-ray and Infrared Fluxes From SN~1987A\tablenotemark{1}}
\tablehead{
\colhead{day\tablenotemark{2}} &
 \colhead{ X-ray flux\tablenotemark{3}} &
 \colhead{ IR flux\tablenotemark{4}} &
 \colhead{ $IRX$\tablenotemark{5}} 
  }
 \startdata 
6067 &  $(1.21\pm0.18)\times10^{-12}$  &  $(3.15\pm0.46)\times10^{-12}$ &      $2.6\pm0.5 $\\
6184 &  $(1.40\pm0.19)\times10^{-12}$  &  $(5.38\pm0.37)\times10^{-12}$ &      $3.8\pm0.6 $\\
6552 &  $(2.72\pm1.0)\times10^{-12}$  &  $(7.43\pm0.38)\times10^{-12}$  &      $2.7\pm1.0$ \\
6734 &  $(3.47\pm0.9)\times10^{-12}$  &  $(8.52\pm0.40)\times10^{-12}$  &     $ 2.4\pm0.7$ \\
6829 &  $(3.62\pm0.79)\times10^{-12}$  &  $(9.07\pm0.38)\times10^{-12}$  &     $ 2.5\pm0.6$ \\
7159 &  $(4.43\pm0.66)\times10^{-12}$  &  $(1.12\pm0.04)\times10^{-11}$  &      $2.5\pm 0.4$ \\
7310 &  $(4.84\pm0.90)\times10^{-12}$ &  $(1.22\pm0.04)\times10^{-11}$  &      $2.5\pm 0.5 $\\
7490 &  $(5.10\pm0.65)\times10^{-12}$  &  $(1.32\pm0.04)\times10^{-11}$  &     $ 2.6\pm0.3$ \\
7690 &  $(5.83\pm1.18)\times10^{-12}$  &  $(1.43\pm0.04)\times10^{-11}$ &      $2.5\pm0.5$ \\
7983 &  $(5.79\pm1.00)\times10^{-12}$  &  $(1.55\pm0.04)\times10^{-11}$  &     $ 2.7 \pm 0.5$ \\
 \enddata
  \tablenotetext{1}{All fluxes are in units of erg~cm$^{-2}$~s$^{-1}$.}
 \tablenotetext{2}{Measured since the explosion.}
\tablenotetext{3}{Soft X-ray flux in the 0.5-2.0~keV band, interpolated to the epochs of the \spitz/MIPS observations and corrected for an extinction column density of $N_H = 2.35\times10^{21}$~cm$^{-2}$.}
\tablenotetext{4}{IR fluxes on days 6184 to 7983 were derived from the MIPS 24~\mic\ data using the conversion factor of 1.64 (see Fig. \ref{IRspec_vol}). The flux on day 6067 was derived using the 10.4~\mic\ T-ReCS flux as a proxy for the integrated 5-30~\mic\ spectrum. }
  \tablenotetext{5}{The ratio of  the IR to soft X-ray flux from the ER.}
  \label{table:IRXvol}
\end{deluxetable*}

\begin{figure*}[ht] 
 \centering
   \includegraphics[width=3.0in]{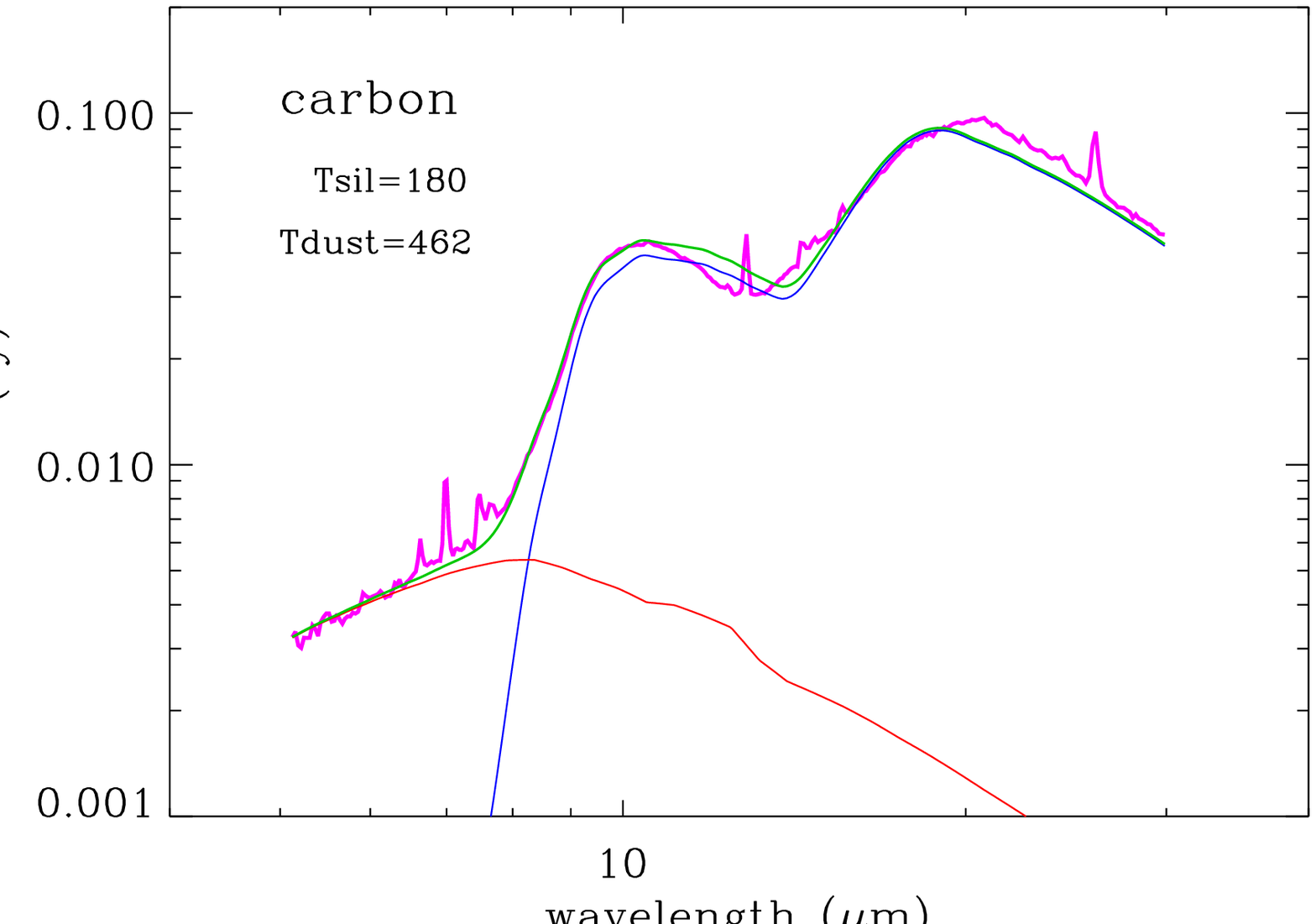}
  \includegraphics[width=3.0in]{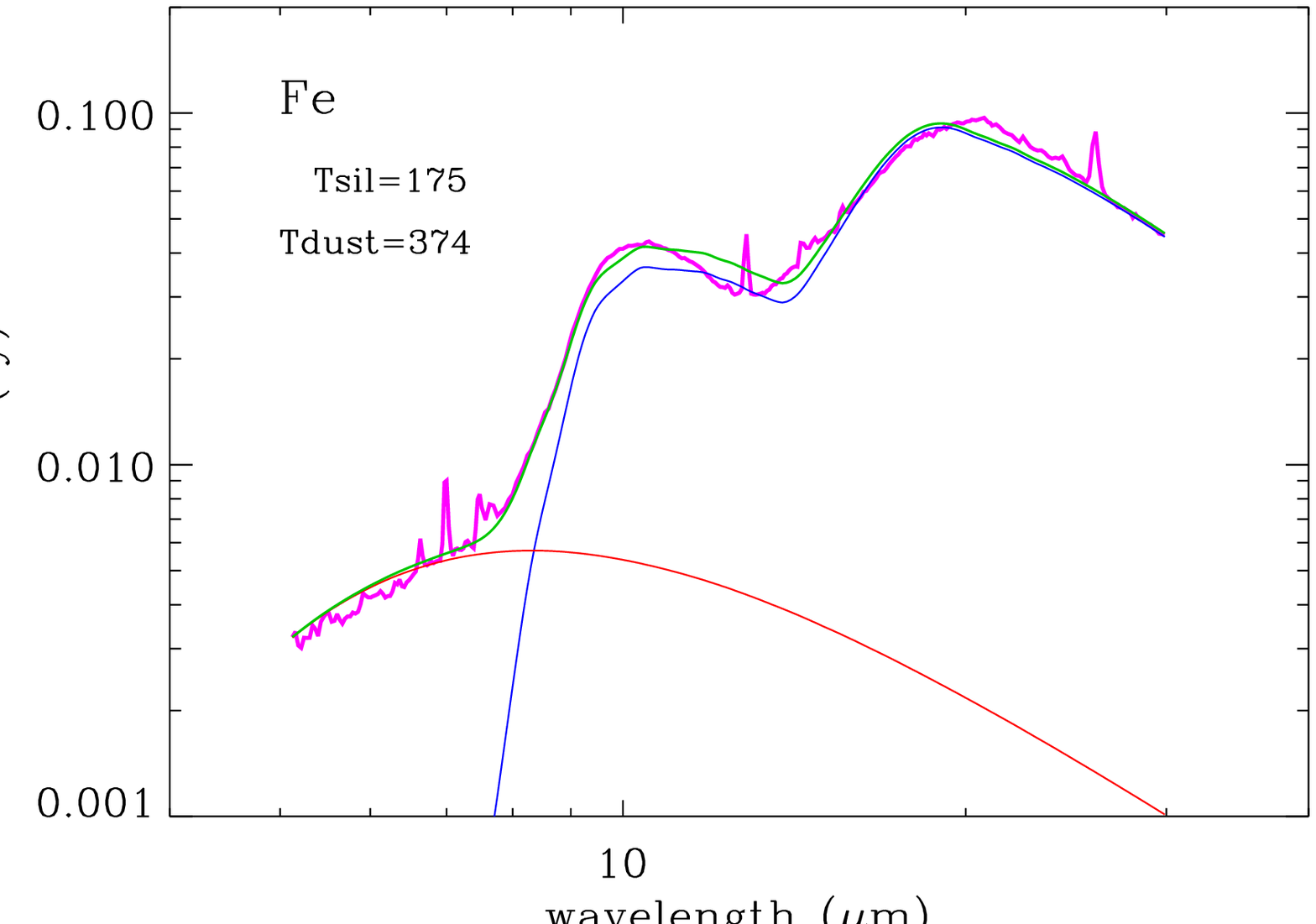}\\
  \includegraphics[width=3.0in]{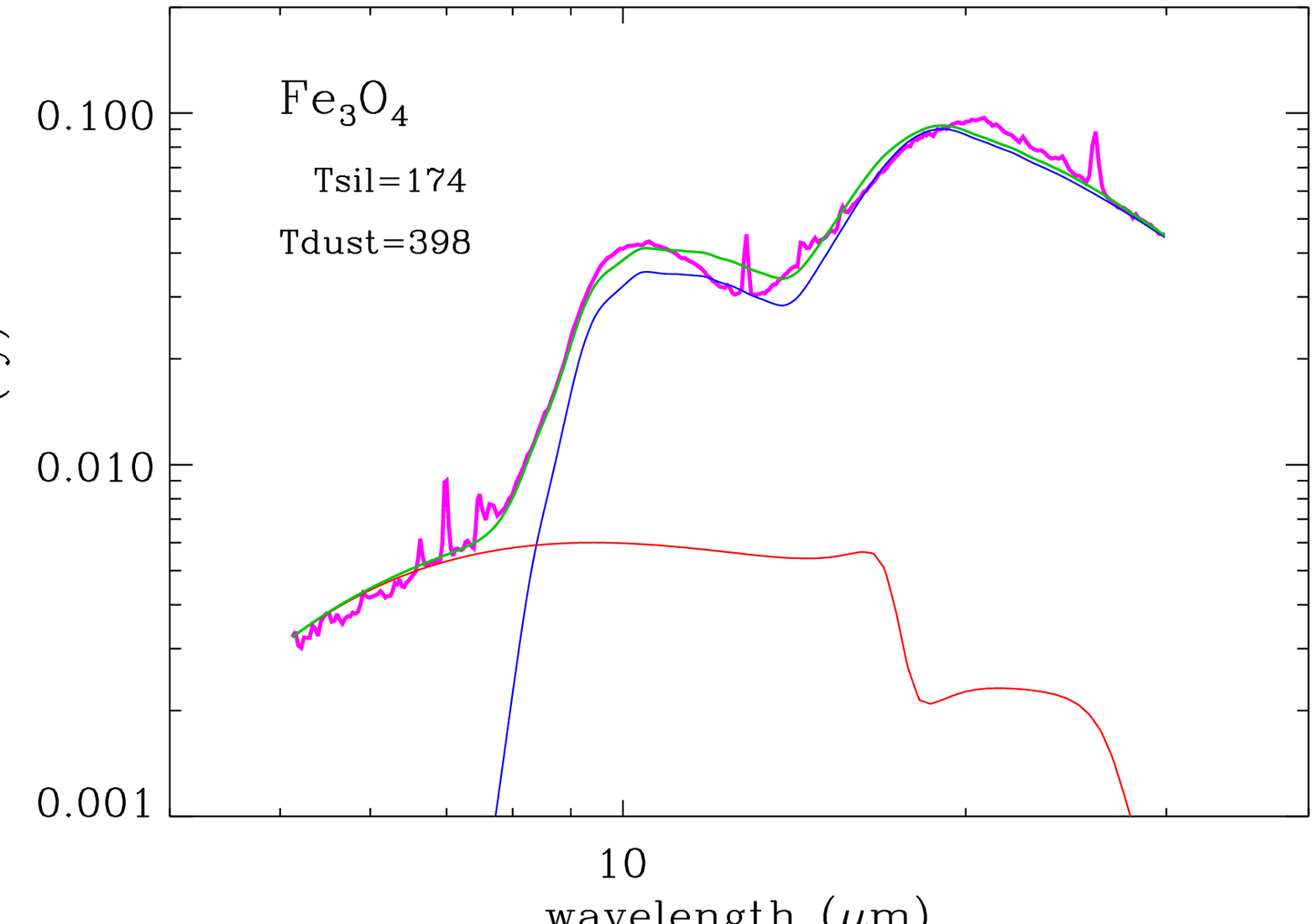}
  \includegraphics[width=3.0in]{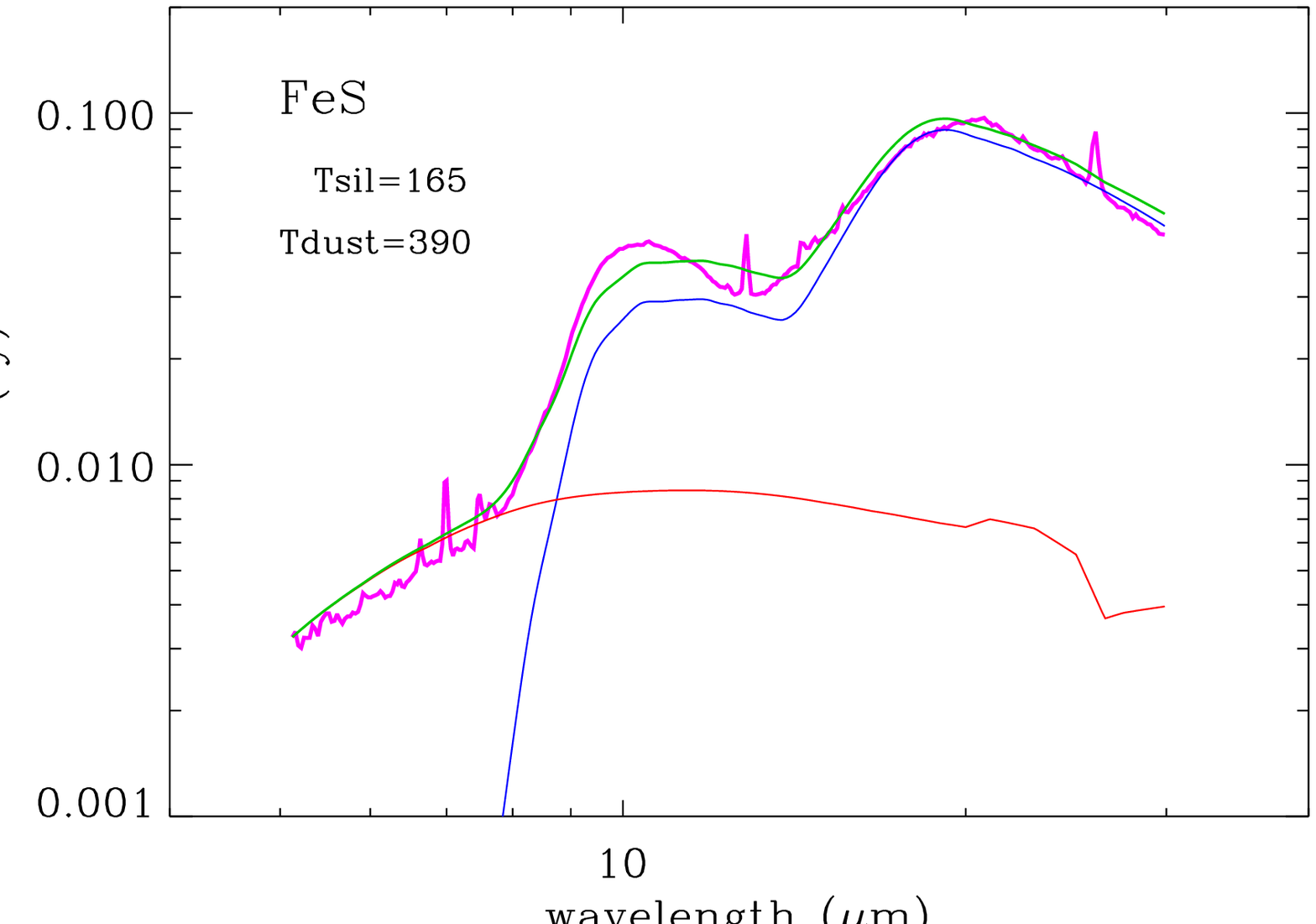}\\
  \caption{\footnotesize{Two-composition fits to the observed infrared spectrum of the ER, indicated by the violet curve. An additional hot component consisting of either carbon, metallic iron, iron oxide, or iron sulfite grains (red curves) provides a good fit to the $5-8$~\mic\ segment of the spectrum. The blue curve represents the spectrum of the silicate grains, and the green curve represents the sum of the emission from the silicate and the secondary dust component. }}
  \label{sil_comp}
\end{figure*} 

\section{THE EVOLUTION OF THE INFRARED EMISSION AND THE DUST COMPOSITION IN THE ER}

\begin{figure*}[ht] 
 \centering
 \includegraphics[width=4.0in]{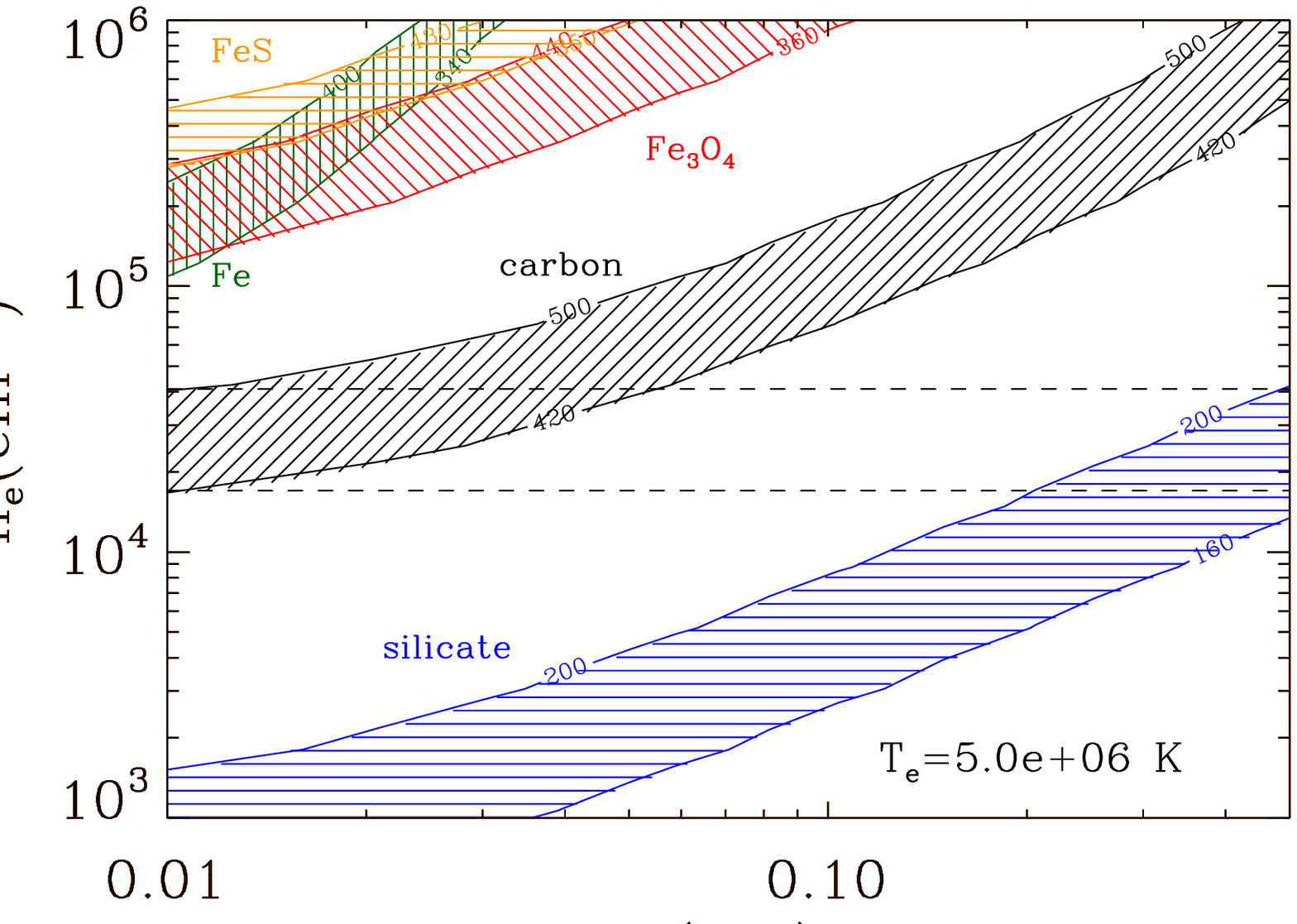} 
 \vspace{0.1in} 
  \caption{\footnotesize{The density required to collisionally heat the different grain species to their respective temperatures given in Figure \ref{sil_comp} as a function of grain radius. Details in text.}}
  \label{contour}
\end{figure*} 

Figure \ref{IRspec_vol} depicts the evolution of the IRAC and MIPS broad band photometry (top left panel), and the low-resolution IRS spectrum of SNR~1987A (top right panel), as a function of time. 
The relatively strong background may affect the continuum at $\lambda > 20$~\mic\ and the [Ne~III] 15.6 $\micron$ and [S~III] 18.7, 33.5~\mic\ emission lines. Line emission from SN 1987A and its surrounding medium will be presented elsewhere (Arendt et al; in preparation). 
 The figures show that the IR flux is rising smoothly in all bands, increasing by a factor of $\sim 3$ between days 6190 and 7980.  The bottom left panel of the figure shows the spectra normalized by least-square fit to the one obtained on day 7554. The spectrum maintained a nearly constant shape throughout the 1800 days of observations. This constancy enables the use of any IRAC or 24~\mic\ MIPS band as a proxy of the total integrated IR flux from the ER. The bottom right panel in the figure shows the ratio between the integrated IR flux and the 24~\mic\ flux from the ER. The ratio is constant with a value of 1.64 to within $<$~10\%. The constancy in the spectral shape also places strong constraints on the parameters of the hot X-ray gas in which the grains are embedded. 

Figure \ref{sil_bb} (left panel) shows that the mean spectrum of the ER can be well fitted by astronomical silicate grains radiating at a single temperature of $\sim 180$~K. The relative ratio between the two peaks of the silicate features provide strong constraints on the temperature. The mass of the silicate dust is $1.2\times10^{-6}$~\msun\ on day 7554, and should scale linearly with the IR flux for other epochs (see Table 1). The figure also shows the presence of an IR excess at wavelengths between $\sim 5-8$~\mic\ that cannot be attributed to emission from the silicate grains. This excess emission was first detected by \cite{bouchet06}, but was not further pursued because of the shorter integration time and larger noise of that early \spitz\ spectrum. An unidentified K-band (2.2~\mic) continuum at a level of $\sim 0.4$~mJy, detected by \cite{kjaer07}, may be part of this emission component. This emission component, now clearly present in all epochs, can be approximated in the 5 to 8~\mic\ wavelength region by a $\nu^{\alpha}$ power law with a spectral index of $\alpha \approx -1.8$.   This rules out thermal bremsstrahlung, which has a well defined spectral index of $\alpha \approx -0.1$ in this frequency regime, as a possible source. Synchrotron radiation can also be ruled as a source of this emission component. The observed radio synchrotron emission \citep{zanardo10} has a spectral index of $\approx -0.8$, and its extrapolation from a value of 300~mJy at 1.4~GHz on day~7000 will make a negligible contribution to the 8~\mic\ emission around that epoch. 
We therefore conclude that this emission arises from an additional hot dust component that is tightly correlated with the silicate emission. Its spectral shape should be featureless at longer wavelengths as to not reduce the goodness of the fit to the observed 9.7 and 18~\mic\ silicate emission features.  The right panel of the figure shows the fit of a blackbody spectrum, modified by a $\lambda^{-2}$ emissivity law and a temperature of $\sim 350$~K, to this component. A $\lambda^{-2}$ emissivity is consistent with metallic or carbon grain composition, rather than dielectric or silicate composition \citep{bohren83}.

\begin{figure*}[ht] 
\centering
  \includegraphics[width=5.0in]{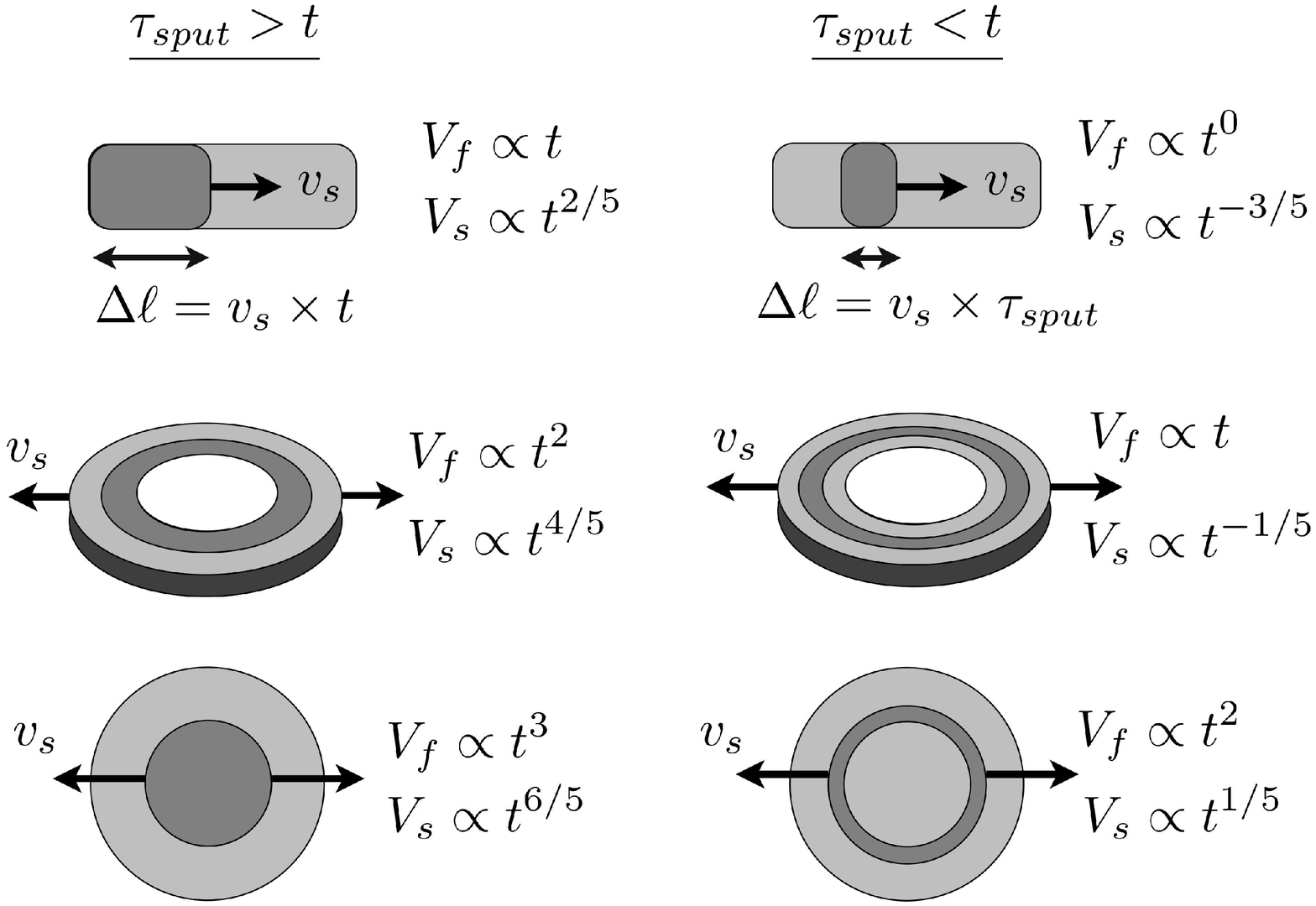}
  \caption{\footnotesize{A schematic view of the evolution of the volume of shocked swept-up dust for different geometries of the circumstellar medium: a linear protrusion (top row); a hollow disk (middle row); a uniform sphere (bottom row). The amount of radiating hot dust is depicted as a dark-shaded region behind the advancing shock. The left column depicts the evolution when grain destruction is insignificant, i.e. when $\tau_{sput} > t$, where $t$ is the time when the shock first encounters the circumstellar medium. The mass of shock-heated dust increases in proportion to the swept up gas. The right column  depicts the volume of the shock-heated dust when $\tau_{sput} > t$. The amount of radiating dust is then constant as the increase in the mass of the swept-up dust is balanced by its destruction. Each scenario exhibits a different dependence on time, which also depends on the expansion law of the blast wave: $V_f$ is the volume swept up for a freely-expanding blast wave ($v_s \sim t^0$), and $V_s$ is the volume swept if the blast wave entered the Sedov phase of its evolution ($v_s \sim t^{-3/5}$).}}
  \label{cartoon}
\end{figure*} 

The modified blackbody is of course an idealized representation of this secondary dust component. Figure \ref{sil_comp} shows a two-component fit to the IR spectrum when the secondary component was taken to consist of either carbon, pure iron, magnetite (Fe$_3$O$_4$), or iron sulfide (FeS) grains. All components, combined with the main silicate dust component, provided acceptable fits to the overall IR spectrum of the ER. The temperature of the secondary dust component is significantly higher than that of the silicate grains, ranging from $\sim 370$~K for pure iron grains to $\sim 460$~K for carbon grains. Some other possible secondary dust components, in particular iron- and aluminum-oxides, were ruled out because these grain species produce strong emission features at wavelengths longward of $\sim 8$~\mic\ that are not present in the \spitz\ IRS spectra.  

The problem of accounting for the widely different temperatures of the two dust components, even though they may be collisionally-heated by the same gas, will be addressed in \S4 below.

\begin{figure}[ht] 
  \includegraphics[width=3.5in]{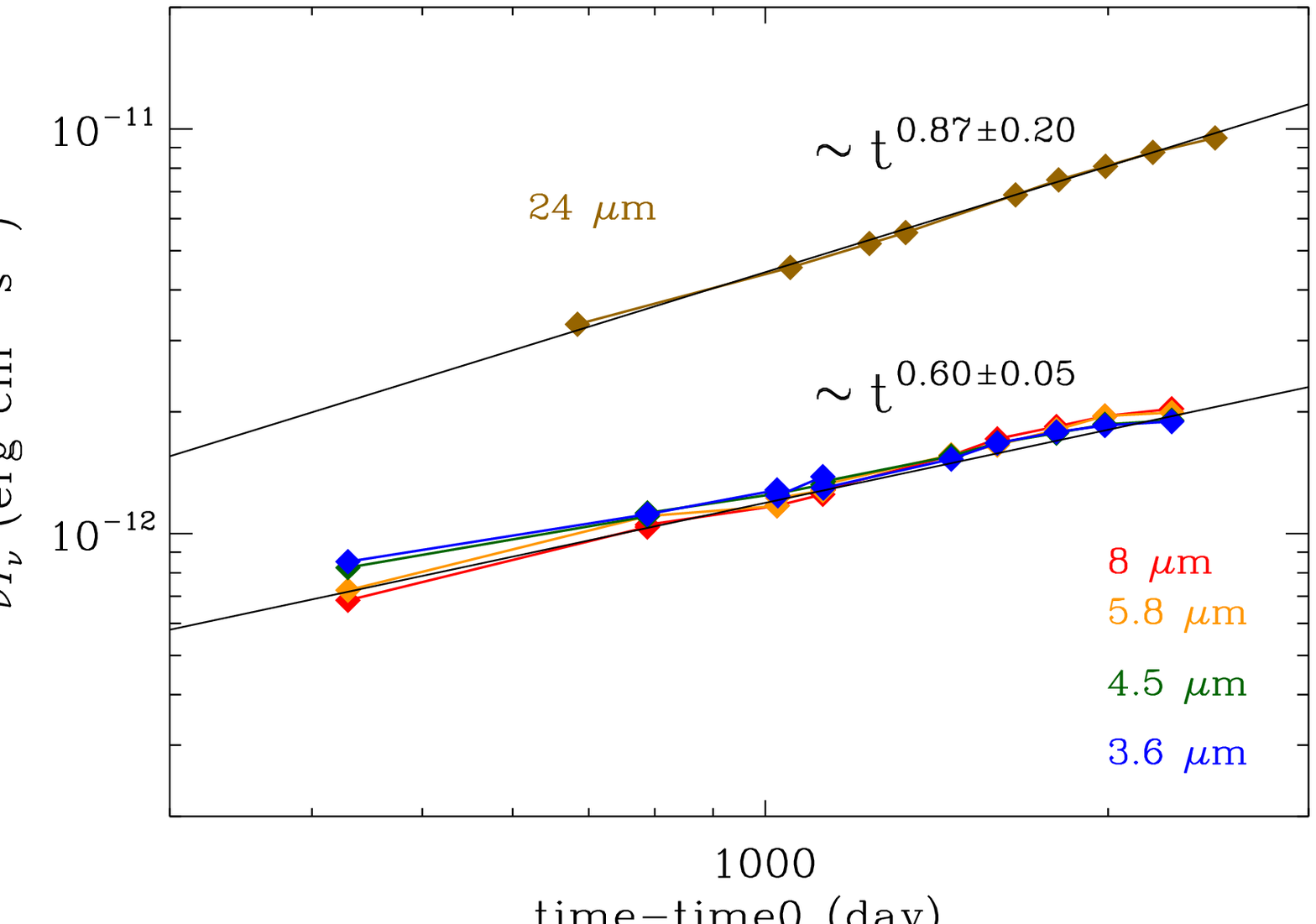}
  \caption{\footnotesize{The evolution of the (renormalized) IRAC and MIPS 24~\mic\ intensities as a function of $t'\equiv t-t_0$, where $t_0$ is the time at which the SN blast wave first encountered the ER. The value of $t_0$ is equal to 5700~d for the IRAC fluxes, and equal to 5500~d for the 24~\mic\ flux. The total IR flux emitted by the silicate dust (represented by the evolution of the 24~\mic\ luminosity) exhibits a steeper rise than that of the secondary dust component, represented by the IRAC luminosities. The rise in the IR spectrum is consistent with a Sedov blast wave expanding into a disk-like medium, with negligible grain destruction see Figure \ref{cartoon}.}}
  \label{IRAC_vol}
\end{figure}

\section{PLASMA CONDITIONS AS DERIVED FROM IR OBSERVATIONS}

Figure \ref{contour} shows the plasma densities required to collisionally heat the different grain types to their respective temperatures given in Figure \ref{sil_comp} as a function of grain radius. The grains are assumed to be embedded in an ionized plasma with a temperature of $5\times10^6$~K, reflecting the average temperature of the soft X-ray component (see \S5 below). If we require both dust components to arise from the same gas, then only the carbon grains support a range of plasma densities (enclosed between the two horizontal dashed lines) that overlap with that of the silicate grains. So the IR spectrum of the ER is consistent with the emission from a population of small ($a \lesssim 0.05$~\mic) carbon grains intermixed with a population of large ($a \gtrsim 0.20$~\mic) silicate grains embedded in a plasma with a density $n_e \approx n_H \approx (2-4)\times 10^4$~\cc. Requiring a common environment for the two dust components rules out the other dust compositions (Fe, Fe$_3$O$_4$, and FeS) as viable dust candidates. Alternatively, the emission could arise from these non-carbonaceous grains, provided they reside in a significantly denser phase ($n_e \gtrsim 2\times 10^5$~\cc) of the X-ray emitting gas. This phase then has to be tightly correlated with the lower-density one in order to explain the tight correlation between the IR light curves of the two dust emission components.

In principle, the second dust component could consist of grains that have formed in the SN ejecta. There is ample evidence for the formation of dust in SN1987A \citep{moseley89b,lucy91,wooden93}. In particular, the disappearance of IR fine structure line emission from Fe has been interpreted as evidence for the formation of iron dust in the ejecta \citep{dwek92c,wooden97}. However, it is unlikely that this second dust component is ejecta dust. The total IR luminosity of this second dust component is about $10^{36}$~erg~s$^{-1}$, calculated for a distance of 50~kpc to the SN. This luminosity is higher by a factor of $\sim 10^2$ from that expected from the radioactive decay of $^{44}$Ti, assuming that $\sim10^{-4}$~\msun\ of $^{44}$Ca was produced in the explosion \citep{woosley89}. So the second dust component can only be heated by a reverse shock expanding through the ejecta. 
As shown below, the IR light curve depends on the rate at which the mass of the dust is swept up and its rate of destruction. These rates depend on the medium morphology and density, and the shock velocity. The tight correlation between the light curves is unlikely to arise from separate shocks propagating into vastly different environments such as the ejecta and the ER. It is therefore very unlikely that the emission from the second dust component arises from SN-condensed dust in the ejecta.        

The potential presence of carbon with the silicate dust in the ER is surprising, defying the common paradigm that either carbonaceous or silicate dust can form in the wind of the progenitor, depending on the C/O abundance ratio in the outflow. The observations therefore suggest either that CO formation did not exhaust all the carbon in the outflow, or that the co-existence of the two dust components is a manifestation of a binary origin of the ER. The progenitor of SN~1987A could have shared a common envelope with a less massive carbon-rich star. The interaction between them could have led to the formation of the ER as well as the two outer rings observed by the Hubble to be in the vicinity of the SN \citep{morris09}.

\begin{figure*}[ht] 
 \centering
\includegraphics[width=3.0in]{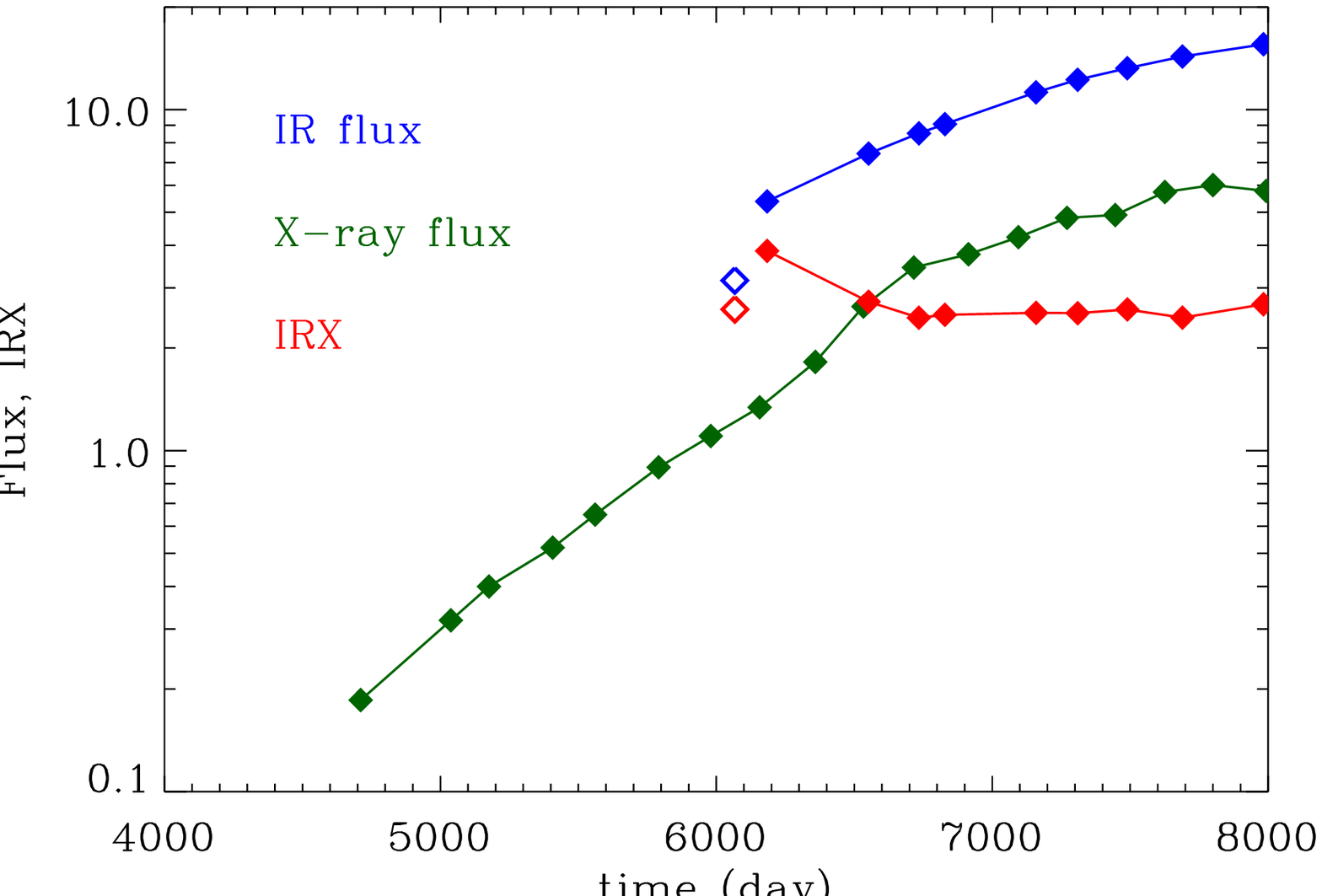}
  \includegraphics[width=3.0in]{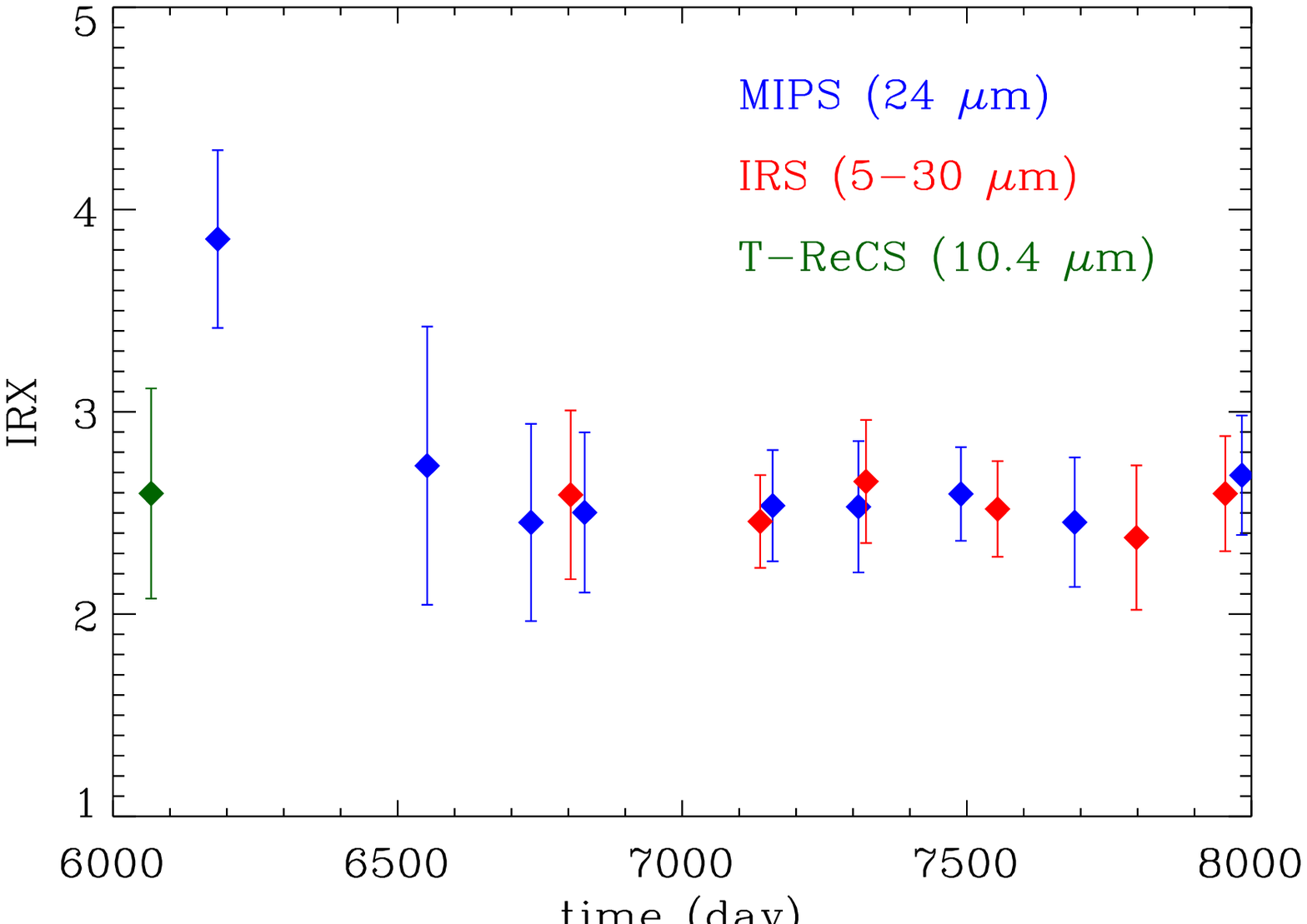}
  \caption{\footnotesize{{\bf Left panel}: The evolution of the IR flux, represented by the 24~\mic\ MIPS (filled blue diamonds) and 10.4~\mic\ T-ReCS (open blue diamond) observations, the soft X-ray flux (derived from the \chan\ observations), and IRX, the IR-to-X-ray flux ratio. Infrared and X-ray fluxes are given in units of 10$^{-12}$~erg s$^{-1}$~cm$^{-2}$. IRX is dimensionless. {\bf Right panel}: Details of the evolution of IRX. Solid blue diamonds and the open green diamond are derived using the MIPS 24~\mic\ and the T-ReCS 10.4~\mic\ fluxes as proxies for the integrated IR spectrum. The IRX derived using the integrated IR fluxes is shown as filled red diamonds. The data are consistent with a constant value of IRX over the entire $\sim 6000 - 8000$~d epoch.}}
  \label{IRX_vol}
\end{figure*} 

\section{THE EVOLUTION OF THE INFRARED AND X-RAY EMISSION AND THEIR RATIO}
A rise in the IR emission is expected as the shock sweeps up more dust during its propagation into the dusty circumstellar medium.
The rate of increase of the IR emission  depends on the morphology of the medium into which the shock is expanding, the expansion rate of the shock, and on the possibility that dust may be destroyed in the postshock gas. The expected increase of the IR emission with time is schematically illustrated in Figure~\ref{cartoon}. The left column represents the case in which the grain lifetime is sufficiently long that it can be ignored, that is, $\tau_{sput} > t'$, where $t'$ is the residence time of the dust in the hot gas. The right column, for which $\tau_{sput} <t' $, represents the case in which the grain lifetime is sufficiently short that an equilibrium is established between the rate of grain destruction and the rate at which the dust is replenished by the expanding shock. For both cases the figure depicts a shock expanding into  a 1-dimensional protrusion (top), a 2-D disk, or ring (middle), and a homogeneous sphere (bottom). The dependence of the volume of the swept-up gas is presented for two different phases in the evolution of a blast wave: (1) the free expansion phase, during which the velocity, $v_f$, is constant; and (2) the Sedov-Taylor phase, which commences after most of the kinetic energy of the explosion has thermalized, during which the velocity $v_s \sim t'^{-3/5}$ \citep{zeldovich67}.

When $\tau_{sput} > t'$, following the left column of Figure~\ref{cartoon}, the intensity of the IR emission scales as the mass of swept up dust, which for a constant dust-to-gas mass ratio, simply scales as the volume of the shocked gas. For a freely-expanding blast wave, the volume of the shocked gas, $V_f$, will increase as $t'$, $t'^2$, and $t'^3$, for the three morphological scenarios, respectively. During the Sedov phase the volume of the shocked gas, $V_s$, will be increasing of increasing more slowly as $t'^{2/5}$, $t'^{4/5}$, and $t'^{6/5}$, respectively. 

When $\tau_{sput} < t'$, following the right column of Figure~\ref{cartoon}, the dust column density through the shock will eventually reach a constant value, roughly equal to $v_s\times \tau_{sput}$, where $\tau_{sput}$ is the grain destruction time in the postshock gas. The IR intensity behind a freely-expanding shock will therefore reach a constant value if the shock is expanding into a one dimensional protrusion, and will increase as $t'$ or $t'^2$ when it expands into a ring or sphere, respectively. During the Sedov phase the volume of the shocked gas, $V_s$, will actually be decreasing as $t'^{-3/5}$, $t'^{-1/5}$, or increasing more slowly as $t'^{1/5}$, for the three morphological scenarios, respectively.

Figure \ref{IRAC_vol} depicts the evolution of the IRAC and 24~\mic\ MIPS fluxes. The fluxes were fitted by a power-law fit of the form $(t-t_0)^{\alpha}$, where $t$ is the time since the explosion, and $\alpha$ and $t_0$ are free parameters. The value of $t_0$ can be interpreted as the time when the shock first encountered the ring. The values of \{$\alpha, t_0$\} were found to be \{0.6, 5700~d\} and \{0.87, 5500~d\} for the IRAC and MIPS fluxes, respectively. 

The dust flux scales as the volume, and the value of $\alpha$ derived from the light curve is consistent with $V_s \sim t'^{4/5}$ (left column of Figure~\ref{cartoon}), indicating that grain destruction is unimportant. More specifically, the 24~\mic\ flux increases at a rate that is consistent with an Sedov-Taylor blast wave expanding into a circumstellar disk/ring. As shown below, the inferred dust lifetime is consistent with currently used values for the sputtering rates of dust grains in hot plasmas.

The value of $t_0$ is consistent with the time inferred from X-ray observations that showed a deceleration of the radial expansion of the SN around day 6000 \citep{racusin09}, and the upturn in the soft X-ray luminosity that occurred around day 6200 \citep{park06}.  The IR and X-ray observations therefore offer a consistent scenario in which both fluxes originate from the dust and gas shocked by the blast wave that has started to penetrate the main body of the ER around day 6000 after the explosion.

\begin{figure*}[ht] 
 \centering
 \includegraphics[width=4.0in]{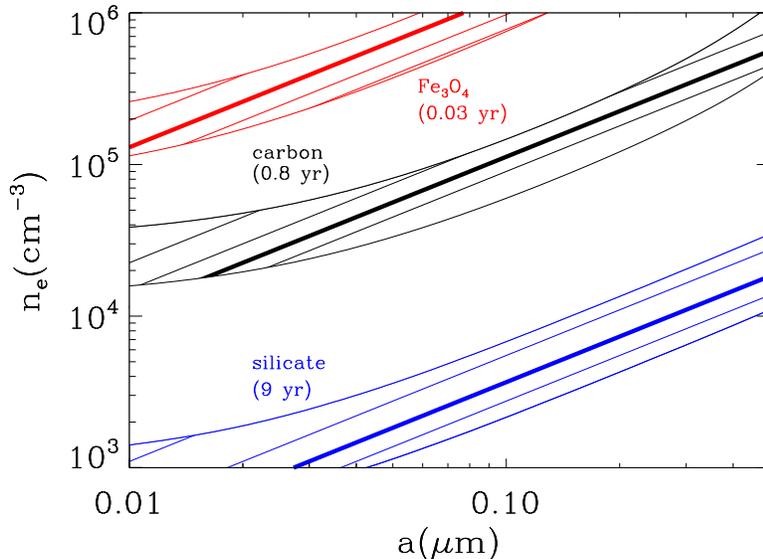} 
 \vspace{0.1in} 
  \caption{\footnotesize{The sputtering lifetime of the different grain species calculated for the same gas density--grain radii combinations required to heat the different dust species to their observed temperature. Dust species are color coded as in Figure \ref{contour}.  
  Within the permitted range of silicate grains, the straight lines indicate
the loci of constant sputtering lifetimes $\tau_{sput}$ = 3, 6, 9 (bold), 12 yr.
For carbon dust the lines indicate $\tau_{sput}$ = 0.4, 0.6, 0.8 (bold), 1.0 yr.
For iron oxide dust the lines indicate $\tau_{sput}$ = 0.02, 0.03 (bold), 0.04, 0.05 yr.}}
  \label{sput}
\end{figure*}

\subsection{The Evolution of $IRX$}
Analysis of the \chan\ X-ray observations of the ER show that the spectrum is generated by two distinct emission components: a soft component with temperatures $\sim (0.3-0.6)$~keV, and a harder component characterized by temperatures between $\sim (2-5)$~keV \citep{zhekov09, park06}. The emission from the soft component arises from the gas heated by the shock that is transmitted into the ER. The hard component arises from the lower density gas interior to the ER that is shocked twice: once by the advancing SN blast wave, and a second time by the shock reflected from the ER. We assume here that the IR emission arises from the soft X-ray component generated by the shock that is penetrating the ER.

\begin{figure*}[ht] 
 \centering
 \includegraphics[width=3.0in]{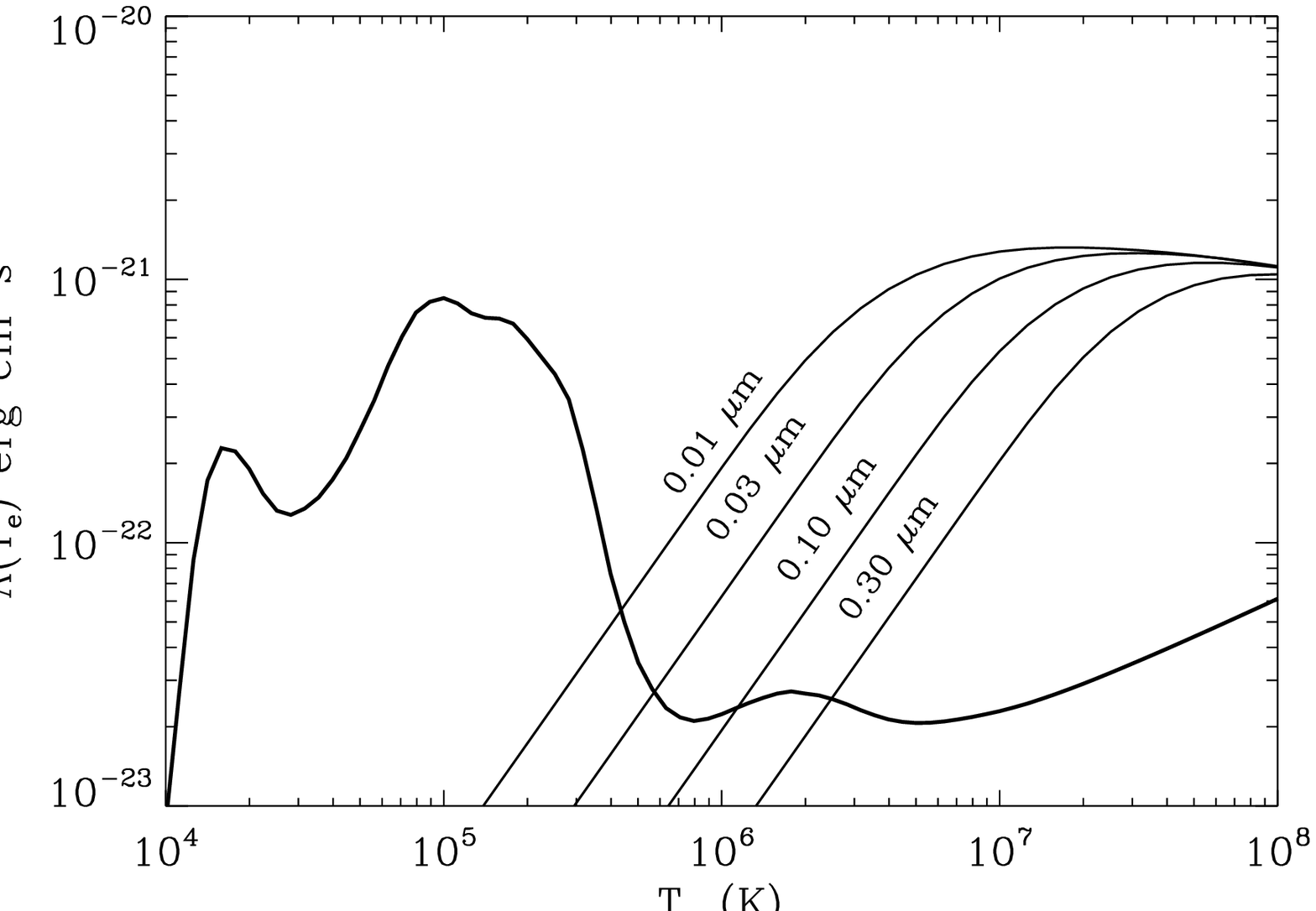} 
    \includegraphics[width=3.0in]{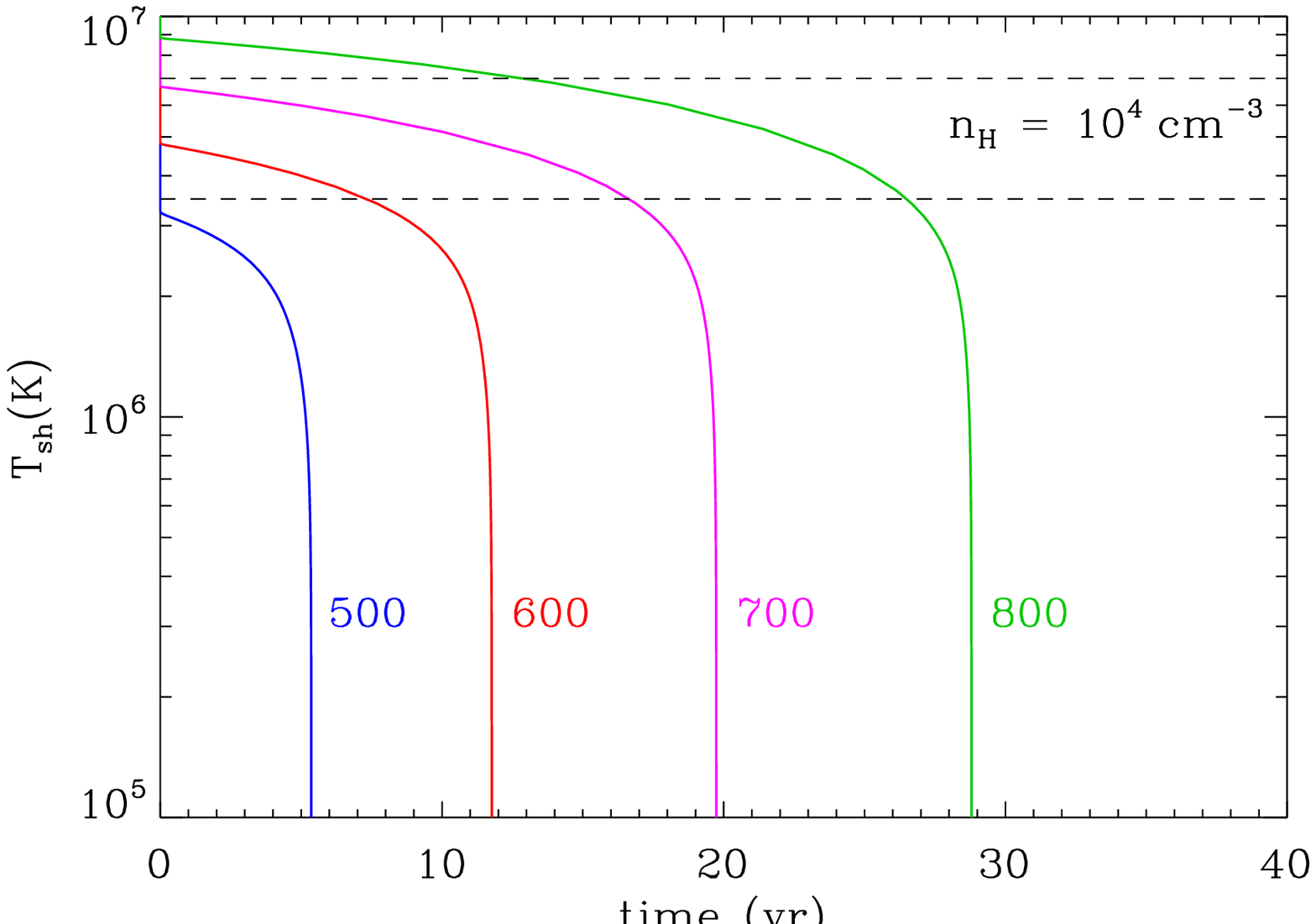} 
  \caption{\footnotesize{{\bf Left panel}: The equilibrium cooling curve for a plasma with ER abundances (thick line). The thin lines represent the gas cooling via dust-gas collisions. All curves assume that all the dust is represented by silicate grains with radii indicated in the figure. {\bf Right panel}: The evolution of the temperature behind shocks of different velocities, expanding through a homogeneous dusty medium with ER abundances and a preshock hydrogen density of $1\times10^4$~\cc.}}
  \label{coolrate}
\end{figure*} 

Figure \ref{IRX_vol} (left panel) depicts the evolution of the soft X-rays and IR fluxes, and their ratio, \irx, as a function of time. The IRS observations only cover the $\sim 6804 - 8000$~d epoch, so we used the MIPS 24~\mic\ observations that cover the additional $\sim 6190-6800$~d epoch as a proxy for the integrated 5-30~\mic\ emission. The figure also depicts the day 6067 datum point, for which we used the scaled mid-IR photometry from Gemini T-ReCS 10.4~\mic\ observations \citep{bouchet06} as a proxy for the integrated IR flux. 

The radial expansion of the soft X-ray component slowed down around day $\sim 6000$, which was interpreted as evidence for the transition of the blast wave from the free-expansion to the Sedov-Taylor phase as it enters the main body of the ER \citep{park05,racusin09}. The T-ReCS datum suggests that this transition has also manifested itself in the IR emission, re-enforcing the consistency between the X-ray and IR evolution of the remnant. 

The evolution of \irx\ is shown in more detail in the right panel of the figure. It has a roughly constant value of $\sim 2.5$ throughout the entire $\sim 6000-8000$~d epoch. An exception is the 6190 datum point. We believe that the anomalously large value of \irx\ at this point is caused by a relative dip in the X-ray flux starting near day 5800 and ending with the rise in the soft X-rays near day 6200. The large value of \irx, also suggests that a significant fraction of the refractory elements in the ER are locked up in dust.

A constant value of \irx\ indicates that the IR and X-ray fluxes evolve at the same rate. This requires the dust-to-gas mass ratio to be uniform throughout the medium into which the shock is expanding, and that neither grain destruction nor gas cooling affect the IR and X-ray emission from the shocked gas. Alternatively, the constancy in the ratio could be the result of both processes, grain destruction and gas cooling, being important. In this case, the IR and X-ray emission have attained a steady state in which the losses to grain destruction and cooling are balanced by the influx of new material such that the volume of radiating gas and dust remains constant in time.
 However, the rate of increase in the IR fluxes suggest that the volume of the shock heated dust is increasing, so that the second scenario is a-priori not likely to be the cause for the constancy in \irx. In the following we examine in more detail the physical condition of the shocked gas that can lead to a constant \irx.

\subsection{Grain Destruction in the Hot Gas}
The lifetime of a dust grain of radius $a$ moving at a high velocity through a hot gas with temperatures  between $\sim 10^6 - 10^7$~K is given by \citep{nozawa06,dwek08a}:
\begin{equation}
\label{tausput}
\tau_{sput}(yr) = A_{sput}\, {a(\mu m)\over n_H(cm^{-3})}
\end{equation}
where $A_{sput} = 3.3\times 10^5, 9.0\times 10^5,\, 5.6\times 10^5,\, 2.8\times 10^5$,and $3.9\times10^5$, respectively, for silicate, carbon, iron, FeS, and Fe$_3$O$_4$ dust, calculated for grains moving at  velocities $\gtrsim 500$~km~s$^{-1}$ through a $5\times 10^6$~K gas. The values of the velocity and temperature characterize those of the shock giving rise to the soft X-ray component. 
Figure \ref{sput} depicts the sputtering lifetime of the dust as a function of grain radii and gas density.   The figure shows that the lifetime of the silicate grains is between $\sim$~4 and $\sim$~15 yr, depending on the combination of grain size and gas density. In contrast, for the secondary dust component this same range of gas temperatures and densities yields significantly shorter grain lifetimes, between $\sim 0.4$ and $\sim 1$~yr, which are too short to explain the observed temporal evolution of the secondary dust component. 
 
\subsection{Cooling of a Shock-Heated Dusty Plasma}
Figure \ref{coolrate} depicts the equilibrium atomic cooling rate, $\Lambda(T_e)$, of a gas with ER composition (thick line, left panel) as a function of electron temperature, $T_e$. The ER abundances were taken from \cite{mattila10}, and were derived from AAT and VLT observations of the supernova between days $\sim 1400$ and $\sim 5000$ after the explosion. The thin lines in the figure show the cooling of the plasma via gas-grain collisions, assuming a pure silicate dust composition. The silicate dust-to-gas mass ratio was assumed to be 0.0029, half the solar value. The different curves assume that all the dust is represented by grains with the indicated radii. Dust cooling dominates the cooling via atomic transitions at temperatures above $\sim 10^6$~K, depending on grain size. The value of \irx\ is the ratio of these two cooling rates. From the fact that the initial value of \irx\ is only $\sim 2.5$ we can conclude that the grain sizes in the ER must be relatively large, with typical radii of $\gtrsim 0.2$~\mic, consistent with the sizes required to heat the dust to a temperature of $\sim 180$~K (see Fig. \ref{contour}). 
Using this lower limit on the grain radii in eq. (\ref{tausput}), gives a postshock density of about $(2-4)\times 10^4$~\cc, where we assumed an upper limit of 0.5~\mic\ on the grain radii. 

The equilibrium cooling may not represent the actual cooling of a shocked gas, which may be far from ionization equilibrium. The right panel in Figure \ref{coolrate} depicts the temperature profile behind shocks of different velocities expanding into a homogeneous dusty gas with a preshock hydrogen density of $10^4$~\cc\ and ER composition. The shock models include the effects on non-equilibrium ionization (NEI), and the effect of dust cooling. The $\sim 600-700$~km~s$^{-1}$ shocks well represent the $3-7$~keV temperature of the soft X-ray emitting plasma \citep{park06,zhekov09}. The figure shows that the cooling time of this shocked gas is about $12-20$ yr for a preshock gas density of $n_H = 10^4$~\cc. The IR observations suggest that the postshock density is $\sim (2-4)\times 10^4$~\cc, giving a gas cooling time of $\sim 20-40$~yr. 

\subsection{Scenarios for the Constancy of $IRX$}
The results presented above show that the timescales for grain destruction and gas cooling are sufficiently long that neither processes has significantly affected the IR or X-ray emissions, or their flux ratio, \irx. The sputtering timescale suggests that grain destruction may become important only at day $\sim 6000 + 3200 \approx 9200$ after the explosion, which is about 25~yr. The cooling time suggests that the X-ray emission may not be affected until $t \sim 20-40$~yr. So the scenario in which both the cooling time and the sputtering time are longer than the age of the shocked gas, is a viable explanation for the constancy of \irx. 

The alternative scenario (see \S5.1) suggests that the approximate constancy in \irx\ is the result of both processes, grain destruction and gas cooling, having attained a steady-state, in which the amount of preshocked gas and dust entering the shock is balancing the amount of dust that is destroyed and the gas that has cooled. Such scenario requires the grain destruction timescale and gas cooling time to be significantly shorter than $\sim 100$~d, requiring postshock densities that are more than an order of magnitude larger than those inferred from the temperature of the collisionally-heated silicate dust. 

The original scenario in which the evolution of the IR and X-ray emission from the postshock gas has not yet been affected by the two processes offers therefore a more reasonable explanation for the constancy of \irx. It relies on the existence of a population of relatively large, $a \gtrsim 0.2$~\mic, silicate grains, embedded in a gas with postshock densities of about $(2-4)\times 10^4$~\cc. These numbers should be regarded as suggestive values for the model parameters. 

\subsection{The Mystery of the Secondary Grain Component}
However, this simple explanation is complicated by the presence of the secondary dust component consisting of much smaller dust grains.  Figure \ref{sput} shows that the density-radius combinations required to heat the carbon dust to $\sim 460$~K will also destroy the grains in less than half a year, and the various iron bearing grains would last for less than $\sim 10$~d, which definitely rules them out as viable candidates for the secondary dust component.

The short sputtering time for the carbon dust suggests that their column density through the shock reached a steady-state between  injection and destruction shortly after the shock penetrated the ER. Consequently, the IR emission from these dust particles should have increased at a different rate than the emission from the silicate grains. This is in contrast to the good correlation between the two emission components.

Figure \ref{IRAC_vol} shows that the $5-8$ \mic\ emission does increase slightly less rapidly than the 24~\mic\ emission. However, the difference in slope is only 1/4 as large as expected if the $\tau_{sput}({\rm hot\ dust}) < t' < \tau_{sput}$(silicate). It is therefore possible the $\sim 5-8$~\mic\ emission could arise from a yet unknown dust component consisting of grains with either radii or IR emissivities (or both) sufficiently small to reach the observed high temperature. They should also have a featureless spectrum in the $\sim 4-40$~\mic\ wavelength region, in order not to reduce the goodness of the fit to the $\sim 9-20$~\mic\ silicate spectrum. Their binding energies should be sufficiently large, or the ambient density should be sufficiently low, so that their sputtering lifetime be longer than $\sim 5$~yr in spite of their small radius.  The nature of this secondary dust component is still an unresolved aspect of the observations. 
 
\section{SUMMARY}
In this paper we presented the mid-IR evolution of SN 1987A over a 5 year 
period spanning the epochs between days $\sim$ 6000 and 8000 since the explosion. Its radiative output during this epoch is dominated by the interaction of the SN blast wave with the pre-existing equatorial ring (ER). 
The main results of this paper can be briefly summarized as follows:
\begin{enumerate}
\item The $\sim 8-30$~\mic\ mid-IR spectrum is dominated by emission from $\sim 180$~K silicate dust, collisionally-heated by the hot X-ray emitting gas with a temperature and density of $\sim 5\times10^6$~K and $\sim (2-4)\times 10^4$~\cc, respectively. The mass of the radiating dust is $\sim 1.2\times 10^{-6}$~\msun\ on day 7554, and scales linearly with IR flux.
\item A secondary emission component dominates the spectrum in the $\sim 5-8$~\mic\ region. Its intensity and spectral shape rule out any possible gas or synchrotron emission mechanism as the source of this emission. It must therefore attributed to a secondary dust component radiating at temperatures above $\sim 350$~K.   
\item The overall shape of the $\sim 5-40$~\mic\ dust spectrum has not changed during the observations, suggesting that the density and temperature of the soft X-ray emitting gas have not significantly changed during the more than 5 years of IR observations. The constancy in the spectral shape of the IR emission also suggests that the mass ratio of the silicate to the secondary dust component remained roughly constant during this period.
\item The evolution of the IRAC and MIPS fluxes can be described by a power law in time since the first shock-ER encounter. The silicate emission increases as $t^{0.87}$, consistent with X-ray observations, suggesting that the blast wave has transitioned from a free expansion to the Sedov phase, and is now expanding into the main body of the ER.
\item The infrared-to-X-ray flux ratio, \irx, is constant with a value of $\sim 2.5$ throughout this epoch. The magnitude of \irx\ shows that the cooling of the shocked gas is dominated by IR emission from the collisionally-heated dust with radii $\gtrsim 0.2$~\mic, and that a significant fraction of the refractory elements in the ER should be depleted onto dust.
\item The constancy of \irx\ is consistent with the premise that neither grain destruction by sputtering nor cooling of the shocked gas have played a significant role in the evolution of the IR and X-ray emission. This scenario is consistent with the sputtering rates currently used in the literature as expressed in eq. (1).
\item The presence of a secondary dust component, radiating at significantly higher temperatures than the silicate dust, suggests that the grain radii or IR emissivities of this component must be significantly smaller than those of the silicates. Their sputtering lifetime could therefore be significantly shorter than that of the silicate grains and their evolution distinctly different from that of the silicates, especially around the $\sim 6100$~d time interval. However, a significantly shorter grain lifetime contradicts the nearly constant or slightly decreasing flux ratio between the hotter secondary component and the silicate grain component. Hence, the nature of the secondary dust component remains a mystery. 
\end{enumerate}

Five years of continued monitoring of SN1987A have lead us to conclude there is no current evidence for grain destruction. 
This conclusion stands in contrast to the results of our earlier analysis \cite{dwek08a}, which were based only on two epochs of data 
at early times, and which we now see to be somewhat anomalous. Continuing observations will reveal evidence for grain destruction in the ER, and may elucidate the nature of the mysterious secondary dust component.

{\acknowledgements}
We benefitted from useful comments by Dick McCray and Svetozar Zhekov. We also thank the referee Diane Wooden for her careful reading of the manuscript and many constructive suggestions. This work is based in part on observations made with the Spitzer Space Telescope, which is operated by the Jet Propulsion Laboratory, California Institute of Technology under a contract with NASA. Support for this work was provided by NASA.

\bibliography{/Users/edwek/science/00-Bib_Desk/Astro_BIB.bib}


\end {document}